\documentstyle[prd,aps,preprint,psfig]{revtex}

\def\4he{\,{{}^4{\rm He}}}
\def\li7{\,{{}^7{\rm Li}}}

\def\mev{\,{\rm MeV}}
\def\gev{\,{\rm GeV}}
\def\tev{\,{\rm TeV}}
\def\sec{\,{\rm sec}}
\def\mb{\, {\rm mb}}
\newcommand{\order}{{\cal O}}

\begin{document}
%\twocolumn[\hsize\textwidth\columnwidth\hsize\csname
%@twocolumnfalse\endcsname
%%
%%
\tighten
\draft
\title{MeV-scale Reheating Temperature and Thermalization of Neutrino
Background}
\author{M. Kawasaki and  K. Kohri}
\address{Research Center for the Early
Universe, School of Science, University of Tokyo, Tokyo 113-0033,
Japan}
\author{Naoshi Sugiyama}
\address{Department of Physics, Kyoto University, Kyoto 606-8502, Japan}
\date{\today}

\maketitle
\begin{abstract}
     The late-time entropy production by the massive particle decay
     induces the various cosmological effects in the early epoch and
     modify the standard scenario. We investigate the thermalization
     process of the neutrinos after the entropy production by solving
     the Boltzmann equations numerically.  We find that if the large
     entropy are produced at $t \sim 1$ sec, the neutrinos are not
     thermalized very well and do not have the perfect Fermi-Dirac
     distribution. Then the freeze-out value of the neutron to proton
     ratio is altered considerably and the produced light elements,
     especially $\4he$, are drastically changed. Comparing with the
     observational light element abundances, we find that $T_R
     \lesssim 0.7$~MeV is excluded at 95 $\%$ C.L. We also study the
     case in which the massive particle has a decay mode into hadrons.
     Then we find that $T_R$ should be a little higher, {\it i.e.} 
     $T_R \gtrsim$ 2.5 MeV - 4 MeV, for the hadronic branching ratio
     $B_h = 10^{-2}-1$. Possible influence of late-time entropy
     production on the large scale structure formation and temperature
     anisotropies of cosmic microwave background is studied. It is
     expected that the future satellite experiments (MAP and PLANCK)
     to measure anisotropies of cosmic microwave background radiation
     temperature can detect the vestige of the late-time entropy
     production as a modification of the effective number of the
     neutrino species $N_{\nu}^{\rm eff}$.
\end{abstract}


%]

%%%%%%%%%%%%%%%%%%%%%%%%%%%%%%%%%%%%%%%%%%%%%%%%%%%%%%%%%%%%%%%%%%%%%%
\section{Introduction}
%%%%%%%%%%%%%%%%%%%%%%%%%%%%%%%%%%%%%%%%%%%%%%%%%%%%%%%%%%%%%%%%%%%%%%
In the standard big bang cosmology it had been assumed tacitly that
the universe was dominated by the thermal radiation at the early
epoch.  Even in the paradigm of the modern cosmology it is commonly
believed that thermal radiation was produced by the reheating process
after the primordial inflation and they dominated the energy of the
universe at sufficiently early epoch.  Here we ask, ``How early should
the universe be dominated by radiation in order to success the
standard big bang cosmology?''. We could say that the energy of the
universe should be dominated by the radiation at least before the
beginning of the big bang nucleosynthesis (BBN) epoch.  In this paper
we answer the above question.

The various models of the modern particle physics beyond the standard
model predicts a number of unstable massive particles which have long
lifetimes and decays at about BBN epoch. The energy density of the
non-relativistic particles or the oscillation energy density of the
scalar fields (inflaton and so on) decreases as $\rho_{NR}(t) \propto
a(t)^{-3}$, where $a(t)$ is a scale factor. On the other hand since
the radiation energy density decreases more rapidly $\rho(t)\propto
a(t)^{-4}$, if the energy density of the massive non-relativistic
particles or the oscillating scalar fields is large enough, it
immediately dominates the universe as it expands, and the universe
necessarily becomes matter-dominated until the cosmic time reaches to
their lifetime. When the particles decay into standard particles (e.g.
photon and electron), they produce the large entropy and the universe
becomes radiation-dominated again. It is expected that such
process would change the initial condition for the standard big bang
scenario. We call the process ``late-time entropy production''.

Now we have some interesting candidates for late-time entropy
production in models based on supersymmetry (SUSY).  It is known that
gravitino and Polonyi field which exist in local SUSY ({\it i.e.}
supergravity ) theories have masses of $\sim {\cal O}(100{\rm
GeV}-10{\rm TeV})$~\cite{Nilles}. In addition they have long lifetimes
because they interact with the other particle only through gravity.
For example, since Polonyi field~\cite{Polonyi} which has a heavy mass
$\sim 10$~TeV cannot be diluted by the usual inflation, it immediately
dominates the universe and decays at the BBN epoch. Moreover it is also
known that in the superstring theories there exist many light fields
called dilaton and moduli which have similar properties to the Polonyi
field.

Recently Lyth and Stewart~\cite{Lyth} considered a mini-inflation
called ``thermal inflation'' which dilutes the above dangerous scalar
fields. In the thermal inflation scenario, however, the flaton field
which is responsible for the thermal inflation decays at late times.
In particular, if Polonyi (moduli) mass is less than $\sim 1$~GeV
which is predicted in the framework of gauge-mediated SUSY breaking
models~\cite{Giudice}, the sufficient dilution requires that the
flaton decays just before BBN~\cite{Asaka}.  Thus, in thermal
inflation models, one should take care of the late-time entropy
production.

To keep the success of BBN, any long-lived massive particles or the
coherent oscillation of any scalar fields which dominate the universe
at that time must finally decay into the standard particles before the
beginning of BBN. Moreover the decay products would have to be quickly
thermalized through scatterings, annihilations, pair creations and
further decays and make the thermal bath of photon, electron and
neutrinos. Concerning photons and electrons which electromagnetically 
interact, the interaction rate is much more rapid than the
Hubble expansion rate at that time. Therefore it is expected that the
photon and electron which are produced in the decay and subsequent
thermalization processes are efficiently thermalized. The problem is
that neutrinos can interact only through the weak interaction. In the
standard big bang cosmology the neutrinos usually decouple from the
electromagnetic thermal bath at about $T \simeq 2-3$MeV. Therefore it
is approximately inferred that they can not be sufficiently
thermalized at the temperature $T \lesssim $ a few MeV. Namely the
reheating temperature after the entropy production process should be
high enough to thermalize the neutrinos. Though people had ever used
the rough constraints on the reheating temperature between 1MeV -
10MeV, in the previous paper~\cite{kks} we pointed out that the
neutrino thermalization is the most crucial for the successful BBN. In
this paper we describe the detail of the method to obtain the neutrino
spectrum and the formulations to integrate a set of Boltzmann
equations numerically , and we study the constraint on the reheating
temperature using the obtained neutrino spectrum and the full BBN
network calculations with the revised observational light element
abundances.

The above constraint is almost model-independent and hence
conservative because we only assume that the massive particle decay
produces the entropy. However, a more stringent constraint can be
obtained if we assume a decay mode into quarks or gluons. In this case
some modifications are needed for the above description.  When the
high energy quark-antiquark pairs or gluons are emitted, they
immediately fragment into a lot of hadrons (pions , kaons, protons,
neutrons, {\it etc}.). It is expected that they significantly
influence the freeze-out value of neutron to proton ratio at the
beginning of BBN through the strong interaction with the ambient
protons and neutrons.  In the previous paper~\cite{kks} we did not
consider such hadron injection effects on BBN. Therefore we carefully
treat the hadron injection effects in the present paper.

For another constraint, the late-time entropy production may induce
the significant effects on the anisotropies of the cosmic microwave
background radiations (CMB). Lopez {\it et al.}~\cite{Lopez} pointed
out that the CMB anisotropies are very sensitive to the equal time of
matter and radiation. When the reheating temperature is so low that
neutrinos do not be sufficiently thermalized, the radiation density
which consists of photon and neutrinos becomes less than that in the
standard big bang scenario. It may give distinguishable signals in the
CMB anisotropies as a modification of the effective number of neutrino
species $N_{\nu}^{\rm eff}$. With the present angular resolutions and
sensitivities of COBE observation~\cite{COBE} it is impossible to set
a constraint on $N_{\nu}^{\rm eff}$ but it is expected that future
satellite experiments such as MAP~\cite{MAP} and PLANCK~\cite{PLANCK}
will gives us a useful information about $N_{\nu}^{\rm eff}$.  In
addition the above effect may also induce the signals in the observed
power spectrum of the density fluctuation for the large scale
structure as a modification of the epoch of the matter-radiation
equality.

The paper is organized as follows. In Sec.~II we introduce the
formulation of the basic equations and the physical parameters. In
Sec.~III we briefly review the current status of the observational
light element abundances. In Sec.~IV we study the spectra of the
electron neutrino and the mu(tau)-neutrino by numerically solving the
Boltzmann equations, and the constraints from BBN are obtained there.
In Sec~V we investigate the additional effects in the hadron injection
by the massive particle decay.  In Sec.~VI we consider the another
constraints which come from observations for large scale structures
and anisotropies of CMB.  Sec~VII is devoted to conclusions. In
Appendix we introduce the method of the reduction for the nine
dimension integrals into one dimension.

%%%%%%%%%%%%%%%%%%%%%%%%%%%%%%%%%%%%%%%%%%%%%%%%%%%%%%%%%%%%%%%%%%%%%%
\section{Formulation of Neutrino Thermalization}
%%%%%%%%%%%%%%%%%%%%%%%%%%%%%%%%%%%%%%%%%%%%%%%%%%%%%%%%%%%%%%%%%%%%%%
\label{sec:formulation}

%%%%%%%%%%%%%%%%%%%%%%%%%%%%%%%%%%%%%%%%%%%%%%%%%%%%%%%%%%%%%%%%%%%%%%
\subsection{Reheating Temperature}
%%%%%%%%%%%%%%%%%%%%%%%%%%%%%%%%%%%%%%%%%%%%%%%%%%%%%%%%%%%%%%%%%%%%%%
\label{sec:tr}

In order to discuss the late-time entropy production process, we
should formulate the equations which describe the physical process.
Here the reheating temperature $T_{R}$ is an appropriate parameter to
characterize the late-time entropy production. We define the
reheating temperature $T_{R}$ by
\begin{equation}
     \label{eq:3h}
     \Gamma \equiv 3 H(T_{R}),
\end{equation}
where $\Gamma$ is the decay rate (=$\tau^{-1}$) and $H(T_{R})$ is
the Hubble parameter at the decay epoch (t = $\tau$).~\footnote{Since the
  actual decay is not instantaneous, the matter-dominated universe
  smoothly changes into radiation-dominated one. Thus it is rather
  difficult to clearly identify the reheating temperature by observing
  the evolution of the cosmic temperature. Instead we ``define'' the
  reheating temperature formally by Eq.~(\ref{eq:3h})} The Hubble
parameter is expressed by
\begin{equation}
     H = \left(\frac{g_{*}\pi^2}{90}\right)^{1/2}
     \frac{T_{R}^2}{M_{G}},
\end{equation}
where $g_{*}$ is the statistical degrees of freedom for the  massless
particles  and $M_{G}$
is the reduced Plank mass ($= 2.4\times 10^{18}$GeV). Then the
reheating temperature is given by
\begin{equation}
     \label{eq:rtemp}
     T_{R} = 0.554 \sqrt{\Gamma M_{G}}.
\end{equation}
Here we have used $g_{*} = 43/4$. From Eq.~(\ref{eq:rtemp}), we can
see that the reheating temperature has the one to one correspondence
with the lifetime of the parent massive particle.

Here we define the effective number of neutrino species
$N_{\nu}^{\rm eff}$ as a parameter which characterize the time evolution of
the energy density of neutrinos. Here $N_{\nu}^{\rm eff}$ is defined by
\begin{equation}
     \label{eq:n-eff}
     N_{\nu}^{\rm eff}
     \equiv
     \frac{\rho_{\nu_e}+\rho_{\nu_{\mu}}+\rho_{\nu_{\tau}}}{\rho_{\rm
     std}},
\end{equation}
where $\rho_{\rm std}$ is the total neutrino energy density in the
standard big bang model ({\it i.e.}  no late-time entropy production and
three neutrino species).

%%%%%%%%%%%%%%%%%%%%%%%%%%%%%%%%%%%%%%%%%%%%%%%%%%%%%%%%%%%%%%%%%%%%%%
\subsection{Basic Equations}
%%%%%%%%%%%%%%%%%%%%%%%%%%%%%%%%%%%%%%%%%%%%%%%%%%%%%%%%%%%%%%%%%%%%%%
\label{sec:fundamental}

When a massive particle $\phi$ which is responsible for the late-time
entropy production decays, all emitted particles except neutrinos are
quickly thermalized and make a thermal bath with temperature $\sim
T_{R}$. For relatively low reheating temperature $T_{R} \lesssim
10$MeV neutrinos are slowly thermalized. Since in realistic situations
the decay branching ratio into neutrinos is very small, we assume that
neutrinos are produced only through annihilations of electrons and
positrons, {\it i.e.}  $e^{+} + e^{-} \rightarrow \nu_{i} +
\bar{\nu}_{i} ( i= e, \mu, \tau)$. The evolution of the distribution
function $f_{i}$ of the neutrino $\nu_{i}$ is described by the
momentum dependent Boltzmann equation~\cite{bernstein}:
\begin{equation}
     \label{eq:Boltzmann}
     \frac{\partial f_{i}(\mbox{\boldmath $p$},t)}{\partial t}
     - H(t) \mbox{\boldmath $p$}\frac{\partial f_{i}(\mbox{\boldmath
     $p$},t)}{\partial \mbox{\boldmath $p$}}
     = C_{i, \rm coll}
\end{equation}
where the right hand side is the total collision term.~\footnote{
The integrated Boltzmann equation~\cite{BBF} is not adequate in the present
problem.  As we show in Sec.~\ref{sec:neu_nu}, the spectral shape of
the momentum distribution obtained by our scheme is much different
from the equilibrium one. It should be noticed that the integrated
Boltzmann equation assumes that the shape of the momentum distribution
is the same as the equilibrium one. Thus we should solve the momentum
dependent Boltzmann equation.} When the reaction is two bodies
scattering $1 + 2 \rightarrow 3 + 4$, it is given by the expression,
\begin{eqnarray}
     \label{eq:collision}
     C_{i, \rm coll} ={ 1\over 2E_1}\sum \int &&  {d^3 p_2 \over 2E_2 (2\pi)^3}
{d^3 p_3 \over 2E_3 (2\pi)^3}{d^3 p_4 \over 2E_4 (2\pi)^3}
\nonumber \\
  && \times(2\pi)^4\delta^{(4)} (p_1+p_2-p_3-p_4)\Lambda(f_1,f_2,f_3,f_4)
  S\, |M|^2_{12\rightarrow 34},
\end{eqnarray}
where $|M|^2$ is the scattering amplitude summed over spins of all
particles, $S$ is the symmetrization factor which is 1/2 for identical
particles in initial and final states, $\Lambda = f_3 f_4
(1-f_1)(1-f_2)-f_1 f_2 (1-f_3)(1-f_4)$ is the phase space factor
including Pauli blocking of the final states. Then the total collision
term $C_{i, \rm coll}$ is expressed by,
\begin{equation}
     \label{eq:annscat}
     C_{i, \rm coll} =  C_{i, \rm ann} + C_{i, \rm scat},
\end{equation}
where $C_{i,\rm ann}$ is the collision term for annihilation processes
and $C_{i, \rm scat}$ is collision term for elastic scattering
processes.  Here we consider the following processes:
\begin{eqnarray}
     \nu_{i} + \nu_{i}& \leftrightarrow & e^{+} + e^{-},
     \nonumber \\
     \nu_{i} + e^{\pm}& \leftrightarrow & \nu_{i} + e^{\pm}.
     \nonumber
\end{eqnarray}
In this paper we have treated neutrinos as Majorana ones ({\it i.e.},
$\nu = \bar{\nu}$). It should be noted that there are no differences
between Majorana neutrinos and Dirac ones as long as they are
massless, and since the temperature is $\order(\mev)$ at least in this
situation, we could have treated them as if they were massless
particles. The relevant reactions are presented in
Table~\ref{table:Mnue} for $\nu_e$ and Table~\ref{table:Mnumu} for
$\nu_{\mu}$ and $\nu_{\tau}$.~\footnote{Here we neglect the neutrino
self-interactions. It may lead to underestimate the kinetic
equilibrium rate for high reheating temperatures. However, we think
that this effect does not change the results very much. The
interactions between electrons and neutrinos are the most important
because they transfer the energy of the thermal bath to neutrinos. The
self-interactions of the neutrinos cannot increase the energy density
of neutrinos but mainly change their momentum distribution.
Furthermore, the neutrino number densities are much smaller than the
electron number density at low reheating temperature with which we are
concerned. Thus differences caused by the neutrino self interactions
are expected to be small.}

The collision terms are quite complicated and expressed by nine
dimensional integrations over momentum space. However, if we neglect
electron mass and assume that electrons obey the Boltzmann
distribution $e^{-p/T}$, the collision terms are simplified to one
dimensional integration form.~\footnote{The errors due to
neglecting the electron mass is small and the deviation is just a few
percent. We show the reasons as follows. The difference between
Fermi-Dirac and Maxwell-Boltzmann distribution ``$df$'' is less than
one at most $df < 1.0$. The week interaction rate is almost expressed
by $\langle\sigma v\rangle n_e/ H(t)$, where $\langle\sigma v\rangle
\sim G_F^2 m_e^2$ and $n_e$ is an electron number density.  Then the
error is at most estimated by, $\langle\sigma_W v\rangle n_e/
H(t)\times df \lesssim 10^{-2}$ (for $T \lesssim$ 0.5MeV). Therefore
the deviation is a few percent and the neglecting the electron mass
does not change the results. The other methods of the approximation to
reduce the integral from nine to two dimensions in which the electron
mass is not neglected are presented in ref.~\cite{HM,DHS}} Then
$C_{i,\rm ann}$ is given by~\cite{SWO,Kawasaki}
\begin{equation}
     \label{eq:C-ann}
     C_{i, \rm ann} =   - \frac{1}{2\pi^2}\int p'^2_idp'_i
     (\sigma v)_i (f_i(p_i) f_i(p'_i)- f_{eq}(p_i)f_{eq}(p'_i)),
\end{equation}
where $f_{eq} (= 1/(e^{p_i/T} +1))$ is the equilibrium distribution and
$(\sigma v)_i$ is the differential cross sections given by
\begin{eqnarray}
     \label{eq:cross}
     (\sigma v)_{e} & = & \frac{4G^2_F}{9\pi}
     (C_V ^2 + C_A ^2)pp',\\
     (\sigma v)_{\mu,\tau} & = & \frac{4G^2_F}{9\pi}
     (\tilde{C_V}^2 + \tilde{C_A}^2)pp',\\
\end{eqnarray}
where we take $C_V=\frac12 + 2\sin^2\theta_W$, $C_A= \frac12 $,
  $\tilde{C_V}=C_V-1$ ($\tilde{C_A}=C_A-1$) and $\theta_W$ is Weinberg
angle ($\sin^2\theta_W \simeq 0.231$)~\cite{PDG}.

As for elastic scattering processes, $C_{i,scat}$ is also simplified to one
dimensional integration (see Appendix), and it is expressed as
\begin{eqnarray}
     C_{i, scat} & = &  \frac{G_F^2}{2\pi^3}(C_V^2+C_A^2)
     \left[ -\frac{f_i}{p_i^2}
       \left( \int_0^{p_i}dp'_i F_1(p_i,p'_i)(1-f_i(p'_i))
         + \int^{\infty}_{p_i}dp'_i
         F_2(p_i,p'_i)(1-f_i(p'_i)\right)\right.
     \nonumber \\
     \label{eq:C-scat}
      & & \left. + \frac{1-f_i(p_i)}{p_i^2}
       \left(\int_0^{p_i}dp'_i B_1(p_i,p'_i)f_i(p'_i)
         + \int^{\infty}_{p_i}dp'_i B_2(p_i,p'_i)f_i(p'_i)\right)\right],
\end{eqnarray}
where $(C_V^2+C_A^2)$ should be replaced by
$(\tilde{C_V}^2+\tilde{C_A}^2)$ for $i=\mu, \tau$, and the functions
$F_1, F_2, B_1, B_2$ are given by

\begin{eqnarray}
     F_1(p, p') & = & D(p, p') + E(p, p')e^{-p'/T},
     \nonumber\\
     F_2(p, p') & = & D(p', p)e^{(p-p')/T} + E(p, p')e^{-p'/T},
     \nonumber\\
     B_1(p,p')  & = & F_2(p',p), ~~B_2(p,p')   =  F_1(p',p),
\end{eqnarray}
where
\begin{eqnarray}
     D(p, p') & = & 2T^4(p^2 + p'^2 + 2T(p-p')+4T^2),
     \nonumber\\
     E(p, p') & = & - T^2[p^2p'^2 + 2pp'(p+p')T
     \nonumber\\
     & & + 2(p+p')^2T^2 + 4(p+p')T^3 + 8T^4].
\end{eqnarray}

Together with the above Boltzmann equations, we should
solve the energy-momentum  conservation equation in the expanding universe:

\begin{equation}
     \label{energy_conservation}
      \frac{d\rho(t)}{dt} =  - 3 H(t) (\rho(t)+P(t)),
\end{equation}
where $\rho(t)=\rho_{\phi}+\rho_{\gamma}+\rho_{e}+\rho_{\nu}$ is the
total energy density of $\phi$, photon, electron and neutrinos and it
is given by
%%
%\begin{equation}
%    \label{rho_tot}
%    \rho(t) = \rho_{\phi}(t) + \rho_{\gamma}(t) + \rho_{e}(t) +
%                 \rho_{\nu}(t)
%\end{equation}
%%
\begin{equation}
     \label{rho_tot}
    \rho(t) = \rho_{\phi}(t) + {\pi^2 T^4_\gamma\over 15} + {2\over \pi^2} 
\int {dq q^2
      E_e\over \exp {(E_e/T_\gamma)} +1 } + {1\over \pi^2}
     \int dq q^3 f_{\nu_e}(q) + {2\over \pi^2} \int dq q^3 f_{\nu_\mu}(q),
\end{equation}
where $E_e = \sqrt{q^2 + m^2_e}$ is the electron energy.
$P(t) \equiv P_{\gamma}(t)+P_{e^{\pm}}(t)+P_{\nu}(t)$ is the total pressure,
\begin{equation}
     \label{eq:pressure}
     P(t) = {\pi^2 T^4_\gamma\over 45} + {2\over \pi^2} \int {dq q^4
     \over 3 E_e [\exp (E_e/T_\gamma) +1 ]} + {1\over
     3\pi^2} \int dq q^3 f_{\nu_e}(q) + {2\over 3\pi^2} \int dq q^3
     f_{\nu_\mu}(q).
\end{equation}
%%
%\begin{equation}
%    \label{tot_pressure}
%    P(t) = P_{\gamma}(t) + P_{e}(t) + P_{\nu}(t),
%\end{equation}
$H(t)$ is the Hubble parameter,
\begin{equation}
     \label{eq:hubble}
      H(t) = \frac{\dot{a}(t)}{a(t)} = \frac{1}{\sqrt{3}M_G}\sqrt{\rho(t)}.
\end{equation}
The time evolution equation of $\rho_{\phi}$ is given by
\begin{equation}
      \frac{d\rho_{\phi}}{dt}  =  -\Gamma\rho_{\phi} -3H\rho_{\phi}.
     \label{eq:rho_phi}
\end{equation}
Practically we solve the time evolution of the photon temperature
instead of Eq.~(\ref{energy_conservation}),
\begin{eqnarray}
     \label{eq:dTdt}
     \frac{d T_{\gamma}}{d t} & = & -
     \frac{- \rho_{\phi}/\tau_{\phi} +
     4H\rho_{\gamma}+3H(\rho_{e^{\pm}}+P_{e^{\pm}})+4H\rho_{\nu} +
     d\rho_{\nu}/dt }{\partial\rho_{\gamma}/\partial T_{\gamma}|_{a(t)} +
     \partial\rho_{e^{\pm}}/\partial T_{\gamma}|_{a(t)}},
\end{eqnarray}
together with Eqs.~(\ref{eq:Boltzmann}),~(\ref{eq:hubble})
and ~(\ref{eq:rho_phi}).

%%%%%%%%%%%%%%%%%%%%%%%%%%%%%%%%%%%%%%%%%%%%%%%%%%%%%%%%%%%%%%%%%%%%%%
\section{Observational light element abundances}
%%%%%%%%%%%%%%%%%%%%%%%%%%%%%%%%%%%%%%%%%%%%%%%%%%%%%%%%%%%%%%%%%%%%%%
\label{sec:obs}
In this section we briefly show the current status of the
observational light element abundances. Concerning the deuterium
abundance, the primordial D/H is measured in the high redshift QSO
absorption systems. For the most reliable D abundance, we adopt the
following value which is obtained by the clouds at z = 3.572 towards
Q1937-1009 and at z = 2.504 towards Q1009+2956~\cite{BurTyt},
\begin{equation}
     \label{lowd}
     {\rm D/H} = (3.39 \pm 0.25) \times 10^{-5}.
\end{equation}
On the other hand, recently the high deuterium abundance is reported
in relatively low redshift absorption systems at z = 0.701 towards
Q1718+4807~\cite{webb}, ${\rm D/H} = (2.0 \pm 0.5) \times 10^{-4}$.
Another group also observes the clouds independently~\cite{tyt_high}.
However, because they do not have full spectra of the Lyman series,
the analyses would be unreliable. More recently Kirkman {\it et
al.}~\cite{kirkman} observed the quasar absorption systems at z = 2.8
towards Q0130-4021 and they obtain the upper bound, D/H $\lesssim 6.7
\times 10^{-5}$.  Moreover Molaro {\it et al.} reported D/H $\simeq
1.5 \times 10^{-5}$ which was observed in the absorber at z = 3.514
towards APM 08279+5255 although it has the large systematic errors in
the hydrogen column density~\cite{MBCV}. Considering the current
situation, we do not adopt the high deuterium value in this paper.
%Here, for high D
%we adopt the following value,
%%
%\begin{equation}
%    \label{highd}
%    D/H = (2.0 \pm 0.5) \times 10^{-4}.
%\end{equation}
%%

The primordial $^4$He mass fraction $Y_p$ is observed in the low
metalicity extragalactic HII regions. Since $^4$He is produced with
Oxygen in the star, the primordial value is obtained to regress to the
zero metalicity O/H $\rightarrow 0$ for the observational data. Using
the 62 blue compact galaxies (BCG) observations, it was reported that
the primordial $Y$ is relatively `` low'', $Y_p \simeq
0.234$~\cite{OliSkiSte}.  However, recently it is claimed that HeI
stellar absorption is an important effect though it was not included
in the previous analysis~\cite{Izo} properly. They found the
relatively ``high'' primordial value, $Y_p = 0.245 \pm 0.004$. More
recently Fields and Olive~\cite{FieOLi} also reanalyze the data
including the HeI absorption effect and they obtain
\begin{equation}
     \label{FieOLi}
      Y_p=0.238 \pm (0.002)_{stat} \pm (0.005)_{syst},
\end{equation}
where the first error is the statistical uncertainty and the second
error is the systematic one. We adopt the above value as the
observational  $Y_p$.

The primordial $^7$Li/H is observed in the Pop II old halo stars. In
general a halo star whose surface effective temperature is low (the
mass is small), has the deep convective zone. For such a low
temperature star, the primordial $^7$Li is considerably depleted in
the warm interior of the star.  On the other hand for the high
temperature stars ($T_{eff} \gtrsim 5500$K), it is known that the
primordial abundance is not changed and they have a ``plateau''of the
$^7$Li as a function of the effective temperature. In addition, though
it is also known that $^7$Li/H decreases with decreasing Fe/H, $^7$Li
still levels off at lower metalicity, [Fe/H]$\lesssim -1.5$, in the
plateau stars.  We adopt the recent measurements which are observed by
Bonifacio and Molaro~\cite{BonMol}. They observed 41 old halo stars
which have the plateau. We take the additional larger systematic
error, because there may be underestimates in the stellar depletion
and the production by the cosmic ray spallation. Then we obtain
\begin{equation}
     \label{li7}
     {\rm log_{10}(^7Li/H)} =-9.76 \pm (0.012)_{stat} \pm (0.05)_{syst}
     \pm (0.3)_{add}.
\end{equation}

%%%%%%%%%%%%%%%%%%%%%%%%%%%%%%%%%%%%%%%%%%%%%%%%%%%%%%%%%%%%%%%%%%%%%%%
\section{Neutrino Thermalization and BBN}
%%%%%%%%%%%%%%%%%%%%%%%%%%%%%%%%%%%%%%%%%%%%%%%%%%%%%%%%%%%%%%%%%%%%%%
\label{sec:neu_nu}

%%%%%%%%%%%%%%%%%%%%%%%%%%%%%%%%%%%%%%%%%%%%%%%%%%%%%%%%%%%%%%%%%%%%%
\subsection{Time evolution of Neutrino spectrum}
%%%%%%%%%%%%%%%%%%%%%%%%%%%%%%%%%%%%%%%%%%%%%%%%%%%%%%%%%%%%%%%%%%%%%%
\label{sec:neutrino}
{The evolution of the cosmic temperature $T$ is shown in
Fig.~\ref{fig:temp}~(a) for $T_{R}=10$~MeV, and (b) for $T_{R}=2$~MeV.
In Fig.~\ref{fig:temp}~(a), it is seen that the temperature decreases
slowly as $t^{-1/4}$, {\it i.e.}  $a^{-3/8}$ before the decay epoch,
$t \simeq \Gamma^{-1}(\simeq 5\times10^{-2}\sec)$ which corresponds to
$T_{R}=10$~MeV. This is because the actual decay is not instantaneous
and $\phi$ decays into radiation continuously at the rate
$\Gamma$~\cite{Kolb-Turner}. Then The universe is still in M.D. After
the decay epoch $t \gg\Gamma^{-1}$, all $\phi$-particles decay and the
temperature decreases as $a^{-1}$ and $t^{-1/2}$. Then the universe
becomes radiation-dominated epoch.  Since at the temperature $T
\lesssim 0.5$MeV ($t \gtrsim 3 \sec$), electrons and positrons
annihilate into photons $e^+e^- \rightarrow 2 \gamma$, the temperature
is slightly heated. From Fig.~\ref{fig:temp}~(b) we can see that the
temperature decreases as $t^{-1/4}$ until the decay epoch ($t \lesssim
0.1 \sec)$ which corresponds to $T_{R}=2$~MeV.  After the decay epoch,
the temperature decreases as $t^{-1/2}$ (R.D.).  In the actual
computation we take a initial condition that there exists the net
radiation energy density though the universe is in M.D. This
represents the situation that the massive particle necessarily
dominated the universe as it expands. On the other hand even if there
are at first no radiation $\rho_R \simeq 0, {\it i.e.}  T \simeq 0$
which corresponds to the initial condition of the oscillation epoch
after the primordial inflation or thermal inflation, the cosmic
temperature immediately tracks the same curve $t^{-1/4}$ and then
their decay establish the radiation dominated universe $T \propto
t^{-1/2}$. Therefore our treatment is quite a general picture for each
entropy production scenario and it does not depend on whether the net
initial radiation energy exists or not, only if once the unstable
non-relativistic particles dominate the energy density of the
universe.

In Fig.~\ref{fig:rho-nu} we show the evolutions of $\rho_{\nu_e}$ and
$\rho_{\nu_{\mu}}$ (=$\rho_{\nu_{\tau}}$) (a) for $T_{R} = 10$~MeV and
(b)$2$~MeV.  From Fig.~\ref{fig:rho-nu}(a) we can see that if $T_{R} =
10$~MeV, cosmic energy density is as same as the case of standard big
bang cosmology. As shown in Fig.~\ref{fig:rho-nu}(b), however, the
energy density of each neutrino species for $T_{R} = 2$~MeV is smaller
than the case of standard scenario. Since the electron neutrinos
interact with electrons and positrons through both charged and neutral
currents, they are more effectively produced from the thermal bath
than the other neutrinos which have only neutral current interactions.
The final distribution functions $f_e$ and $f_{\mu} (=f_{\tau})$ are
shown in Fig.~\ref{fig:distribution} (a) for $T_{R} = 10$~MeV and (b)
$2$~MeV.  For $T_{R} = 10$~MeV, each neutrino is thermalized well and
the perfect Fermi-Dirac distribution is established. For $T_{R} =
2$~MeV, however, the distributions are not thermal equilibrium
forms.~\footnote{As we noted in Sec.~\ref{sec:formulation}, we must
not use the integrated Boltzmann equation instead of the momentum
dependent Boltzmann equation in the present problem because the former
assumes the equilibrium distribution. To see this, let us define the
ratio $R_E$ for a neutrino species by $R_E =
(\rho_{\nu}/n_{\nu})/(3.151\tilde{T_{\nu}})$, where $\rho_{\nu}$ is
the neutrino energy density, $n_{\nu}$ is the neutrino number density,
$\tilde{T_{\nu}}$ is the effective neutrino temperature which is
defined by the neutrino number density as, $\tilde{T_{\nu}} \equiv
\left( 2\pi^2 /(3 \zeta(3)) n_{\nu} \right)^{1/3}$.  Here both
$\rho_{\nu}$ and $n_{\nu}$ are computed by integrating the neutrino
distribution function which is obtained by solving the momentum
dependent Boltzmann equation.  Approximately $R_E$ represents a ratio
of the mean energy per a neutrino to the thermal equilibrium one. If
the neutrino is in thermal equilibrium, $R_E$ is unity.  In the case
of the integrated Boltzmann equation, because it is assumed that the
shape of the neutrino distribution is the same as the equilibrium one
at any times, $R_E$ is necessarily unity.  On the other hand, in the
case of our scheme, i.e. the momentum dependent Boltzmann equation,
$R_E$ can not be unity. We have computed the ratio $R_E$ in some
representative reheating temperatures for electron neutrino and have
found that they deviated from unity more at the lower reheating
temperatures, $R_E$ = 1.00, 1.03 and 1.50 (for $T_R$ = 10 MeV, 3 MeV
and 1 MeV). Moreover at the lower reheating temperature than 1 MeV,
the deviation is much larger. This result tells us that the neutrino
distribution deviates from the thermal equilibrium shape considerably
at the low reheating temperatures and we should solve the momentum
dependent Boltzmann equation.  $R_E$ has a tendency to increase as the
reheating temperature decreases. This is because neutrinos are
produced by the annihilation of electrons-positron pairs whose mean
energy per one particle is larger than that of neutrinos.}

In Fig.~\ref{fig:tr_nnu} we can see the change of the effective number
of neutrino species $N_{\nu}^{\rm eff}$ as a function of the reheating
temperature $T_{R}$. If $T_{R} \gtrsim 7$ MeV, $N_{\nu}^{\rm eff}$ is
almost equal to three and neutrinos are thermalized very well. We can
regard that it corresponds to the initial condition which has ever
been used for the standard big bang cosmology. On the other hand, if
$T_{R} \lesssim 7$ MeV, $N_{\nu}^{\rm eff}$ becomes smaller than
three.

%%%%%%%%%%%%%%%%%%%%%%%%%%%%%%%%%%%%%%%%%%%%%%%%%%%%%%%%%%%%%%%%%%%%%%
\subsection{Neutrino thermalization and neutron to proton ratio}
%%%%%%%%%%%%%%%%%%%%%%%%%%%%%%%%%%%%%%%%%%%%%%%%%%%%%%%%%%%%%%%%%%%%%%
\label{sec:nu_np}
If the neutrinos are not thermalized sufficiently and do not have the
perfect Fermi-Dirac distribution, {\it i.e.} in this case there is the
deficit of the neutrino distribution due to the low reheating
temperature, it considerably influences the produced light element
abundances. In particular, the abundance of the primordial $^4$He is
drastically changed.  The change of the neutrino distribution function
influences the neutrino energy density and the weak interaction rates
between proton and neutron. At the beginning of BBN (T $\sim$ 1 MeV -
0.1 MeV) the competition between the Hubble expansion rate $H$ and the
weak interaction rates $\Gamma_{n \leftrightarrow p}$ determines the
freeze-out value of neutron to proton ratio $n/p$. After the
freeze-out time, neutrons can change into protons only through the
free decay with the life time $\tau_n $.  Since $^4$He is the most
stable light element and the almost all neutrons are synthesized
into $^4$He, the abundance of the primordial $^4$He is sensitive to
the freeze-out value of neutron to proton ratio.

If the neutrino energy density gets smaller than that of the standard
BBN (SBBN), Hubble expansion rate which is proportional to the square
of the total energy density is also decreased. Then the freeze out
time becomes later and the $\beta$ equilibrium between neutrons and
protons continues for longer time. As a result less neutrons are left.
In this case the predicted $^4$He is less than the prediction of SBBN.
The effect due to the speed-down expansion is approximately estimated
by
\begin{equation}
     \label{speed_down}
     \Delta Y \simeq - 0.1 (- \Delta \rho_{tot}/\rho_{tot}),
\end{equation}
where $Y$ is the mass fraction of $^4$He and $\rho_{tot}$ is the total
energy density of the universe.

Moreover, when the electron neutrino is not thermalized, there is
interesting effect by which more $^4$He are produced. The weak
reaction rates are computed by integrating neutrino distribution
functions which are obtained by solving Boltzmann equations
numerically. Using the neutrino distribution functions, the six weak
interaction rates between neutron and proton are represented by
\begin{eqnarray}
     \label{eq:beta_reac1}
     \Gamma_{n \to p e^- \bar{\nu_e}}& = &
     K\int_0^{Q-m_e}dp_{\nu_e}\left[\sqrt{(p_{\nu_e}-Q)^2-m_e^2}
     (Q-p_{\nu_e})
        \frac{p_{\nu_e}^2}{1+e^{(p_{\nu_e}-Q)/T_{\gamma}}}
     \left(1-f_{\nu_e}(p_{\nu_e})\right) \right], \\
     \label{eq:beta_reac2}
     \Gamma_{ne^+ \to p \bar{\nu_e}} & = &
     K \int_{Q+m_e}^{\infty}dp_{\nu_e}\left[\sqrt{(p_{\nu_e}-Q)^2-m_e^2}
     (p_{\nu_e}-Q)
       \frac{p_{\nu_e}^2}{e^{(p_{\nu_e - Q})/T_{\gamma}}+1}
     \left( 1-f_{\nu_e}(p_{\nu_e}) \right) \right], \\
     \label{eq:beta_reac3}
     \Gamma_{n \nu_e \to p e^-} & = &
     K \int_0^{\infty}dp_{\nu_e}\left[\sqrt{(p_{\nu_e}+Q)^2-m_e^2}
     (p_{\nu_e}+Q)
       \frac{p_{\nu_e}^2}{1+e^{-(p_{\nu_e}+Q)/T_{\gamma}}}
     f_{\nu_e}(p_{\nu_e})\right], \\
     \label{eq:beta_reac4}
     \Gamma_{pe^-\bar{\nu_e} \to n} & = &
     K \int_{0}^{Q-m_e}dp_{\nu_e}\left[\sqrt{(p_{\nu_e}-Q)^2-m_e^2}
     (Q-p_{\nu_e})
       \frac{p_{\nu_e}^2}{e^{-(p_{\nu_e - Q})/T_{\gamma}}+1}
     f_{\nu_e}(p_{\nu_e}) \right], \\
     \Gamma_{pe^- \to n \nu_e} & = &
     \label{eq:beta_reac5}
     K \int_{0}^{\infty}dp_{\nu_e}\left[\sqrt{(p_{\nu_e}+Q)^2-m_e^2}
     (Q+p_{\nu_e})
       \frac{p_{\nu_e}^2}{e^{(p_{\nu_e + Q})/T_{\gamma}}+1}
     \left( 1-f_{\nu_e}(p_{\nu_e}) \right)\right], \\
     \Gamma_{p\bar{\nu_e} \to ne^+} & = &
     \label{eq:beta_reac6}
     K \int_{Q+m_e}^{\infty}dp_{\nu_e}\left[\sqrt{(p_{\nu_e}-Q)^2-m_e^2}
     (Q-p_{\nu_e})
       \frac{p_{\nu_e}^2}{1+e^{-(p_{\nu_e - Q})/T_{\gamma}}}
     f_{\nu_e}(p_{\nu_e}) \right],
\end{eqnarray}
where $Q = m_n - m_p = 1.29$ MeV and $K$ is a normalization factor
which is determined by the neutron life time $\tau_n $ as $ K \simeq
(1.636 \tau_n)^{-1}$ and $\tau_n $ is obtained by the
experiments~\cite{PDG}. From the above equations we can see that if
neutrino and anti-neutrino distribution functions are decreased, both
$\beta$ decay rates $\Gamma_{n \rightarrow p} = \Gamma_{n \to p e^-
\bar{\nu_e}}+\Gamma_{ne^+ \to p \bar{\nu_e}}+\Gamma_{n \nu_e \to p
e^-}$ and $\Gamma_{p \rightarrow n} = \Gamma_{pe^-\bar{\nu_e} \to n} +
\Gamma_{pe^- \to n \nu_e}+ \Gamma_{p\bar{\nu_e} \to ne^+}$ are
simultaneously decreased by the following reasons. The dominant
effects by the deficit of the distribution functions are to
decrease the rates $\Gamma_{n \nu_e \to p e^-}$,
$\Gamma_{pe^-\bar{\nu_e} \to n}$ and $\Gamma_{p\bar{\nu_e} \to ne^+}$
which have the neutrino or anti-neutrino in the initial state.  On the
other hand, though
the other rates $\Gamma_{n \to p e^- \bar{\nu_e}}$, $\Gamma_{ne^+ \to
p \bar{\nu_e}}$ and $\Gamma_{pe^- \to n \nu_e}$ which have the
neutrino or anti-neutrino in the final state are slightly increased
due to Fermi-blocking factor $(1-f_{\nu})$, the ratio of the
difference $\Delta f_{\nu}$ to $(1-f_{\nu})$ is much smaller than that
of $\Delta f_{\nu}$ to $f_{\nu}$, {\it i.e.}
\begin{equation}
     \label{diff_dist}
     |\Delta f_{\nu}/(1-f_{\nu})| \ll |\Delta f_{\nu}/f_{\nu}|
     \qquad {\rm for } \  f_{\nu} \ll 1.
\end{equation}
Therefore, the enhancement is small and the latter effect is not
important. In total, both weak interaction rates $\Gamma_{n
\rightarrow p}$ and $\Gamma_{p \rightarrow n}$ are decreased and
become smaller than those of SBBN. In Fig.~\ref{fig:weak_rate} the
weak interaction rates $\Gamma_{n \rightarrow p}$ and $\Gamma_{p
\rightarrow n}$ are plotted. The solid lines denote the case of $T_R =
10$ MeV which corresponds to the standard big bang scenario. The
dotted lines denote the case of $T_R = 1$ MeV. In the plot we can see
that the insufficient thermalization of the neutrino distributions
derives the changes of the weak interaction rates.

The decrease of weak interaction rates gives significant effects on
the abundance of $\4he$. When the weak interaction rate
$\Gamma_{n\leftrightarrow p}$ decreases, Hubble expansion rate becomes
more rapid than that of the weak interaction rate earlier. Namely the
freeze-out time becomes earlier. Then the freeze-out value of neutron
to proton ratio becomes larger than in SBBN and it is expected that
the predicted $^4$He abundance becomes larger.
%Second when the
%interaction rates $\Gamma_{n \to p}$ at which neutrons are changed
%into protons become smaller, less neutrons can turn into protons after
%the freeze-out time.
The above effect is approximately estimated by
\begin{equation}
     \label{dY_gamma}
     \Delta Y \simeq + 0.19 (- \Delta\Gamma_{n\leftrightarrow
     p}/\Gamma_{n\leftrightarrow p})
\end{equation}

In Fig.~\ref{fig:npratio} we plot the time evolution of the neutron to
proton ratio. In Fig.~\ref{fig:npratio}(a) we change only the number
of neutrino species in SBBN. The dotted line denoted the curve of
$N_{\nu}^{\rm eff}=1.37$ which corresponds to the effective number of
neutrino species in the case of $T_R = 2$ MeV in the late-time entropy
production scenario. Then we find that the predicted $n/p$ curve is
lower than that of $N_{\nu}^{\rm eff}=3$ due to only the speed down
effects or the later decoupling. In Fig.~\ref{fig:npratio}(b) we plot
the time evolution of $n/p$ when we change the reheating temperature
in the late-time entropy production scenario. The dotted line denotes
the case of $T_R = 2$ MeV. Comparing to the case of $N_{\nu}^{\rm
eff}=1.37$ in Fig.~\ref{fig:npratio}(a), the $n/p$ ratio becomes
larger. It is because the weak interaction rates are decreased by the
deficit of the distribution function. Moreover in the case of $T_R =
1$ MeV the $n/p$ ratio becomes much larger.

%%%%%%%%%%%%%%%%%%%%%%%%%%%%%%%%%%%%%%%%%%%%%%%%%%%%%%%%%%%%%%%%%%%%%%
\subsection{Neutrino thermalization and light element abundances}
%%%%%%%%%%%%%%%%%%%%%%%%%%%%%%%%%%%%%%%%%%%%%%%%%%%%%%%%%%%%%%%%%%%%%%
\label{sec:nu_BBN}
Next we perform Monte Carlo simulation and the maximum likelihood
analysis~\cite{HKKM} to discuss how the theoretical predictions with
the low reheating temperature scenario agree with the observational
light element abundances.

In Fig.~\ref{fig:pred_he4} we plotted $^4$He mass fraction Y as a
function of $T_R$ at $\eta = 5 \times 10^{-10}$ (solid line). The
dashed line denotes the virtual $^4$He mass fraction computed by
including only the speed down effect due to the change of the
effective number of neutrino species which is shown in
Fig.~\ref{fig:tr_nnu}. The dotted line denotes the predicted value of
Y in SBBN at $\eta = 5 \times 10^{-10}$. For $T_R \gtrsim 7$MeV, the
solid line and dashed line are quite equal to the value in SBBN. As
$T_R$ decreases, both the solid and dashed lines gradually decrease
because of the speed down effect due to the change of $N_{\nu}^{\rm
eff}$. The dashed line continues to decrease as the reheating
temperature decreases.

On the other hand, for T$_R \lesssim 2$ MeV the effect that the weak
interaction rates are weakened due to the deficit of the neutrino
distribution function begins to become important and the predicted
value of $Y$ begins to increase as $T_R$ decreases. For $T_R \lesssim
1$ MeV, since it is too late to produce enough electrons whose mass
is about $m_e$ = 0.511 MeV, the weak interaction rates are still more
weakened and $Y$ steeply increases as $T_R$ decreases.

In Fig.~\ref{fig:eta_tr} we plot the contours of the confidence level
in the $\eta$-$T_R$ plane. The solid line denotes 95
$\%$ C.L. and the dotted line denotes 68 $\%$ C.L. The filled square
is the best fit point between the observation and theoretical
predictions. The observational data are consistent with the high
baryon to photon ratio, $\eta \sim (3-6) \times 10^{-10}$. From
Fig.~\ref{fig:eta_tr} we find that $T_R \lesssim 0.7$~MeV is excluded
at 95 $\%$ C.L.  In other wards $T_R$ as low as $0.7\rm MeV$ is
consistent with BBN. Then $N_{\nu}^{\rm
eff}$ can be as small as $0.1$ and it definitely influences the
formation of the large scale structure and CMB anisotropy as is seen
in Sec.\ref{sec:LSS_CMB}.

%%%%%%%%%%%%%%%%%%%%%%%%%%%%%%%%%%%%%%%%%%%%%%%%%%%%%%%%%%%%%%%%%%%%%%
\section{Hadron injection by massive particle decay}
%%%%%%%%%%%%%%%%%%%%%%%%%%%%%%%%%%%%%%%%%%%%%%%%%%%%%%%%%%%%%%%%%%%%%%
%%%%%%%%%%%%%%%%%%%%%%%%%%%%%%%%%%%%%%%%%%%%%%%%%%%%%%%%%%%%%%%%%%%%%%
\subsection{Hadron Jets and $e^+e^-$ collider experiments}
%%%%%%%%%%%%%%%%%%%%%%%%%%%%%%%%%%%%%%%%%%%%%%%%%%%%%%%%%%%%%%%%%%%%%%
In the previous section we discussed only the case in which the parent
massive particle $\phi$ decays into photons or the other
electro-magnetic particles. In this section we consider the entropy
production process along with the hadron injection, {\it i.e.} the
case in which the massive particle has some decay modes into quarks or
gluons.  Then the emitted quark-antiquark pairs or gluons immediately
fragment into hadron jets and as a result a lot of mesons and baryons,
{\it e.g.}  pions, kaons, nucleons (protons and neutrons ) are emitted
into the electro-magnetic thermal bath which is constituted by photon,
electron and nucleons.

For example, if the gravitino $\psi_{\mu}$ is the parent particle
which produces the large entropy, it could have a hadronic decay mode
({\it e.g.}  $\psi_{\mu} \to \tilde{\gamma} q \bar{q}$ ) with the
branching ratio $B_h \simeq {\cal O}$($\alpha$) at least even if the
main decay mode is only $\psi_{\mu} \to \tilde{\gamma}
\gamma$($\tilde{\gamma}$ : photino)~\cite{DDRG}. Then about 0.6 - 3
hadrons are produced for $m_{\phi} \simeq 1 - 100 \tev$. In addition
the emitted high energy photons whose energy is about $m_{\phi}/2$
scatter off the background photons and could also produce the
quark-antiquark pairs through the electromagnetic interaction. For the
cosmic temperature $\simeq \order{(\mev)}$, the energy in the center
of mass frame is $\sqrt{s} \simeq 2 - 20 \gev$ for $m_{\phi} \simeq 1
- 100 \tev$. Then the number of the produced hadrons is about 2 - 7
which effectively corresponds to the hadron branching ratio $B_h \sim
10^{-2}$ if we assume that the hadron fragmentation is similar to the
results of $e^+e^-$ collider experiments. Thus $B_h$ should not become
less than about $10^{-2}$ for gravitino decay.~\footnote{If the decay
mode $\psi_{\mu} \to \tilde{g} g$ ($\tilde{g}$ : gluino) is
kinematically allowed, the hadronic branching ratio becomes close to
one.}  For the other candidate, if the ``flaton'' is the parent
particle as in thermal inflation model, it would also have a hadronic
decay mode ($\phi \to g g$ )~\cite{Lyth} if the flaton mass is larger
than 1 GeV.

If once such hadrons are emitted to the electro-magnetic thermal bath
in the beginning of BBN epoch (at $T \simeq 10 \mev - 0.1 \mev)$, they
quickly transfer all the kinetic energy into the thermal bath through
the electro-magnetic interaction or the strong interaction.  Through
such thermalization processes the emitted high energy hadrons scatter
off the background particles, and then they induce some effects on
BBN.  Especially, the emitted hadrons extraordinarily inter-convert the
ambient protons and neutrons each other through the strong interaction
even after the freeze-out time of the neutron to proton ratio $n/p$.
For the relatively short lifetime ($\tau_{\phi} \simeq 10^{-2} \sec -
10^2 \sec$) in which we are interested, the above effect induces the
significant change in the previous discussion. Namely protons which
are more abundant than neutrons are changed into neutrons by the
hadron-proton collisions and the ratio $n/p$ increases extremely.
Because $\4he$ is the most sensitive to the freeze out value of $n/p$,
the late-time hadron injection scenario tends to increase $Y_p$.

Reno and Seckel~\cite{RS} investigated the influences of the hadron
injection on the early stage of BBN. They constrained the lifetime of
the parent particle and the number density comparing the theoretical
prediction of the light element abundances with the observational
data. Here we basically follow their treatment and apply it to the
scenario of late-time entropy production with hadron injections.

The emitted hadrons do not scatter off the background nucleons
directly. At first hadrons scatter off the background photons and
electrons because they are much more abundant than nucleons. For $t
\lesssim 200 \sec$, the emitted high energy hadrons are immediately
thermalized through the electro-magnetic scattering and they reach to
the kinetic equilibrium before they interact with the ambient protons
and neutrons. Then we use the threshold cross section $\langle\sigma
v\rangle^{H_i}_{N \rightarrow N'}$ for the strong interaction process
$N + H_i \rightarrow N' + \cdot \cdot \cdot$ between hadron $H_i$ and
the ambient nucleon $N$, where $N$ denotes proton $p$ or neutron $n$.
The strong interaction rate is estimated by
\begin{eqnarray}
     \label{eq:gamma^i_nn}
     \Gamma^{H_i}_{N\rightarrow N'} &=& n_N \langle\sigma v\rangle^{H_i}_{N
     \rightarrow N'}  \nonumber \\
     &\simeq& 10^{8} \sec^{-1} f_N
     \left(\frac{\eta}{10^{-9}}\right)
     \left(\frac{\langle\sigma v\rangle^{H_i}_{N \rightarrow N'}}{10
     \mb} \right)
     \left(\frac{T}{2 \mev}\right)^3,
\end{eqnarray}
where $n_N$ is the number density of the nucleon species $N$, $\eta$
is the baryon to photon ratio ($=n_B/n_{\gamma}$), $n_B$ denotes the
baryon number density ($= n_p + n_n$) and $f_N$ is the nucleon
fraction ($ \equiv n_N/n_B$). This
equation shows that every hadron whose lifetime is longer than ${\cal
O}(10^{-8})$ sec contributes to the inter-converting interaction
between neutron and proton at the beginning of BBN. Hereafter we will
consider only the following long-lived hadrons, (mesons, $\pi^{\pm}$,
$K^{\pm}$ and $K_L$, and baryons, $p$, $\overline{p}$, $n$, and
$\overline{n}$). For the relevant process ( $N + \pi^{\pm}
\to N' \cdot \cdot \cdot$, and $N + K^- \to N' \cdot \cdot \cdot$,
etc.), we can obtain the cross sections in~\cite{RS,ky}. Here we
ignore $K^+$ interaction because $n + K^+ \rightarrow p + K^0$ is the
endothermic reaction which has $Q = 2.8$ MeV.

We estimate the average number of emitted hadron species $H_i$ per
one $\phi$ decay as
\begin{equation}
     \label{eq:NHi}
     N^{H_i} =  B_h N_{jet} f_{H_i} \frac{\langle N_{ch}\rangle}{2},
\end{equation}
where $\langle N_{ch}\rangle$ is the averaged charged-particle
multiplicity which represents the total number of the charged
particles emitted per two hadron jets, $f_{H_i}$ is the number
fraction of the hadron species $H_i$ to all the emitted charged
particles, $B_h$ is the branching ratio of the hadronic decay mode and
$N_{jet}$ is the number of the produced jets per one $\phi$ decay.

Here it is reasonable to assume that the averaged charged particle
multiplicity $\langle N_{ch}\rangle$ is independent of the the source
because the physical mechanism which governs the production of hadron
jets is quite similar and does not depend on the detail of the origin
only if the high energy quark-antiquark pairs or gluons are emitted.
We adopt the data which are obtained by the $e^+e^-$ collider
experiments. LEPII experiments (ALEPH, DELPHI, L3 and OPAL) recently
give us the useful data for $\sqrt{s}$ = 130 $-$ 172 GeV~\cite{PDG}.
We adopt the following fitting function for $\sqrt{s}$ = 1.4 $-$ 172
GeV~\cite{ky},
\begin{equation}
     \label{eq:nch}
     \langle N_{ch}\rangle= 1.73 + 0.268 \exp\left(
    1.42\sqrt{\ln(s / \Lambda^2)}\right),
\end{equation}
where $\sqrt{s}$ denotes the center of mass energy, the functional
shape is motivated by the next-to-leading order perturbative QCD
calculations, $\Lambda$ is the cut-off parameter in the perturbative
calculations and we take $\Lambda = 1$ GeV. In Fig.~\ref{fig:nch} we
plot the charged particle multiplicity for $\sqrt{s} = 1 \gev -
100\tev$.  The error of the fitting is about 10$\%$.  Using the
available data~\cite{PDG,ky,biebel}, we obtain the emitted hadron
fraction $f_{H_i}\equiv n^{H_i}/\langle N_{ch}\rangle$,
\begin{eqnarray}
     \label{eq:fHi}
     f_{\pi^+}=0.64,~~  f_{\pi^-}=0.64,~~\nonumber \\
     f_{K^+}=0.076,~~  f_{K^-}=0.076,~~f_{K_L}=0.054 \\
     f_{p}=f_{\overline{p}}=0.035,~~ f_{n}=f_{\overline{n}}=0.034,
    ~~\nonumber
\end{eqnarray}
where $n^{H_i}$ is the number of the emitted hadron species $H_i$
which is defined as the value after both $K_S$ and $\Lambda^0$ had
completely finished to decay.  \footnote{Although the summation of
$f_{H_i}$ is obviously more than one, it is because the experimental
fitting of $\langle N_{ch}\rangle$ is defined as a value before $K_S$
and $\Lambda^0$ begin to decay~\cite{biebel}. Here we assume that
$f_{H_i}$ do not change significantly in the energy range $\sqrt{s}
\simeq $ 10 GeV - 100 TeV.  Since we do not have any experimental data
for the high energy region more than about 200 GeV, we extrapolate
$\langle N_{ch}\rangle$ to the higher energy regions and we take
$f_{H_i}$ as a constant.} As we find easily, almost all the emitted
particles are pions which are the lightest mesons. To apply the data
of the $e^+e^-$ collider experiments, we take $\sqrt{s}= 2 E_{jet}$ in
$\langle N_{ch}\rangle$ where $E_{jet}$ denotes the energy of one
hadron jet because the $\langle N_{ch}\rangle$ is obtained by the
result for two hadron jets.

%%%%%%%%%%%%%%%%%%%%%%%%%%%%%%%%%%%%%%%%%%%%%%%%%%%%%%%%%%%%%%%%%%%%%%
\subsection{Formulation in Hadron Injection Scenario}
%%%%%%%%%%%%%%%%%%%%%%%%%%%%%%%%%%%%%%%%%%%%%%%%%%%%%%%%%%%%%%%%%%%%%%
In this section we formulate the time evolution equations in the
late-time hadron injection scenario.  As we mentioned in the previous
section, the hadron injection at the beginning of BBN enhances the
inter-converting interactions between neutron and proton equally and
the freeze out value of $n/p$ can be extremely increased. Then the
time evolution equations for the number density of a nucleon $N (=p,
n)$ is represented by
\begin{equation}
     \label{eq:difeqN}
     \frac{dn_N}{dt} + 3 H(t) n_N = \left[\frac{dn_N}{dt}\right]_{weak}
     - \Gamma_{\phi} n_{\phi} \left( K_{N \rightarrow N'} - K_{N'
     \rightarrow N} \right),
\end{equation}
where $H(t)$ is Hubble expansion rate, $[dn_N/dt]_{weak}$ denotes
the contribution from the weak interaction rates which are obtained by
integrating the neutrino distribution functions as discussed in
Sec.~\ref{sec:neu_nu}, see
Eqs.(\ref{eq:beta_reac1} - \ref{eq:beta_reac6}),
$n_{\phi}=\rho_{\phi}/m_{\phi}$ is the number density of $\phi$, $K_{N
\rightarrow N'}$ denotes the average number of the transition $N
\rightarrow N'$ per one $\phi$ decay.

The average number of the transition $N \rightarrow N'$ is expressed by
\begin{equation}
     \label{eq:Knn}
     K_{N \rightarrow N'} = \sum_{H_i} N^{H_i}R^{H_i}_{N \rightarrow N'},
\end{equation}
where $H_i$ runs the hadron species which are relevant to the nucleon
inter-converting reactions, $N^{H_i}$ denotes the average number of
the emitted hadron species $H_i$ per one $\phi$ decay which is given by
Eq.~(\ref{eq:NHi}) and $R^{H_i}_{N \rightarrow N'}$ denotes the
probability that a hadron species $H_i$ induces the nucleon transition
$N \rightarrow N'$,

\begin{equation}
     \label{eq:trans_prob}
     R^{H_i}_{N \rightarrow N'} =
      \frac{\Gamma^{H_i}_{N \rightarrow N'}}{\Gamma^{H_i}_{dec} +
      \Gamma^{H_i}_{abs}},
\end{equation}
where $\Gamma^{H_i}_{dec} = \tau_{H_i}^{-1}$ is the decay rate of the
hadron $H_i$ and  $\Gamma^{H_i}_{abs} \equiv \Gamma^{H_i}_{N \rightarrow
N'}+\Gamma^{H_i}_{N' \rightarrow N}+\Gamma^{H_i}_{N \rightarrow
N}+\Gamma^{H_i}_{N' \rightarrow N'}$ is the total absorption rate of
$H_i$.

%%%%%%%%%%%%%%%%%%%%%%%%%%%%%%%%%%%%%%%%%%%%%%%%%%%%%%%%%%%%%%%%%%%%%%
\subsection{Hadron injection and BBN}
%%%%%%%%%%%%%%%%%%%%%%%%%%%%%%%%%%%%%%%%%%%%%%%%%%%%%%%%%%%%%%%%%%%%%%
In this subsection we compare the theoretical prediction of the light
element abundances in the hadron injection scenario to the
observational light element abundances. In the computations we assume
that the massive particle decays into three bodies
($E_{jet}=m_{\phi}/3$) and two jets are produced at the parton level
($N_{jet}=2$)~\footnote{The above choice of a set of model parameters
$E_{jet}$ and $N_{jet}$ is not unique in general and is obviously
model dependent. However, since $\langle N_{ch}\rangle$ has the
logarithmic dependence of $E_{jet}$, we should not be worried about
the modification of $E_{jet}$ by just a factor of two so seriously.
On the other hand in Eq.~(\ref{eq:difeqN}), the second term in the
right hand side scales as $\propto N_{jet}/m_{\phi}$.  For the
modification of $N_{jet}$, therefore, we only translate the obtained
results according to the above scaling rule and push the
responsibility off onto $m_{\phi}$}.  In the computing we take the
branching ratio of the hadronic decay mode $B_h = \order(10^{-2}- 1)$.

As we noted in the previous subsections, it is a remarkable feature
that the predicted $Y_p$ tends to increase in the hadron injection
scenario because $\4he$ is the most sensitive to the freeze-out value of
the neutron to proton ratio.  Since protons which are more abundant
than neutrons are changed into neutrons through the strong
interactions rapidly, the freeze out value of $n/p$ increase extremely
if once the net hadrons are emitted.  In Fig.~\ref{fig:had_tr_Y} we
plot the predicted $\4he$ mass fraction $Y_p$ as a function of $T_R$
for (a) $m_{\phi}$=100 TeV and (b) $m_{\phi}$ =10 GeV. The solid curve
denotes the predicted $Y_p$.  Here we take the branching ratio of the
hadronic decay mode as $B_h$ = 1 (right one) and $B_h$ = 0.01 (left
one). The dot-dashed line denotes $B_h = 0$. The dashed line denotes
the virtual value of $Y_p$ computed by including only the speed down
effect due to the change of the effective number of neutrino species.
The dotted line denotes the prediction in SBBN.

As we mentioned in the previous section, the speed down effect due to
deficit of the electron neutrino distribution function are not
important for $T_R \gtrsim 7$MeV. In addition since it is high enough
to keep $n/p$ $\simeq$ 1 for the cosmic temperature $T \gtrsim 7$MeV,
the enhancements of the inter-converting interaction between n and p
due to the hadron emission do not induce any changes on the freeze-out
value of $n/p$. As $T_R$ decreases ($T_R \lesssim 7$MeV), $Y_p$ also
decreases gradually because the speed down effect on the freeze-out
value of $n/p$ begins to be important. On the other hand, if a lot of
hadrons are emitted when the cosmic temperature is $T \lesssim$ 6 - 7
MeV and the ratio $n/p$ is less than one, they enhance the
inter-converting interactions more rapidly. As a result, the ratio
$n/p$ attempts to get closer to one again although the cosmic
temperature is still low. Thus the above effects extremely increase
the freeze-out value of $n/p$ and is much more effective than the
speed down effects. Namely the produced $Y_p$ becomes larger very
sensitively only if $T_R$ is just a little lower than 6 - 7 MeV. One
can obviously find that this effect becomes more remarkable for the
larger $B_h$.

To understand how it  depends on mass, it is convenient to introduce the
yield variable $Y_{\phi}$ which is defined by
\begin{equation}
     \label{eq:yield_phi}
     Y_{\phi} \equiv n_{\phi} / s,
\end{equation}
where $s$ denotes the entropy density in the universe. Because
$Y_{\phi}$ is a constant only while the universe expands without any
entropy production, it represents the net number density of $\phi$ per
comoving volume. For the simplicity let's consider
the instantaneous decay of $\phi$ and assume that the reheating
process has been completed quickly. Because the radiation energy in
the thermal bath or entropy $s=2\pi^2g_{*}/45T_R^3$ is produced only
from the decay products of $\phi$, $Y_{\phi}$ is approximately
estimated using $T_R$ and $m_{\phi}$ by
\begin{equation}
     \label{eq:yield2}
     Y_{\phi} \simeq 0.28 \frac{T_R}{m_{\phi}}.
\end{equation}
 From the above equation, we can see that for the fixed value of $T_R$
the net number of $\phi$, {\it i.e.} the net number of the emitted
hadrons, becomes larger for the smaller mass. Comparing
Fig.~\ref{fig:had_tr_Y} (a) with Fig.~\ref{fig:had_tr_Y} (b), we find
that the theoretical curve of $Y_p$ for the case of $m_{\phi}$ = 10
GeV is enhanced more steeply and the starting point to increase $Y_p$
becomes higher than for the case of $m_{\phi}$ = 100 TeV.

Since the other elements (D and $\li7$) are not so sensitive as
$\4he$, it is expected that the observational value of $Y_p$
constrains $T_R$ most strongly. In order to discuss how a low
reheating temperature is allowed by comparing the theoretical
predictions with observational values (D, $\4he$ and $\li7$), we
perform the Monte Carlo simulation and maximum likelihood analysis as
discussed in Sec.~\ref{sec:neu_nu}. In addition to the case of
Sec.~\ref{sec:neu_nu} we take account of the following uncertainties,
the error for the fitting of $\langle N_{ch}\rangle$ as
10$\%$~\cite{ky} and the experimental error for each cross section of
the hadron interaction as 50$\%$. Because there are not any adequate
experimental data for the uncertainties of cross
sections~\cite{RS,ky}, here we take the larger values to get a
conservative constraint.

In Fig.~\ref{fig:m100_eta_tr} we plot the contours of the confidence
level for $m_{\phi}= 100$ TeV in the ($\eta$-$T_R$) plane for (a)
$B_h$ = 1 and (b) $B_h = 10^{-2}$. The solid line denotes 95 $\%$
C.L., the dotted line denotes 68 $\%$ C.L. and the filled square is
the best fit point between the observation and theoretical prediction
for D, $\4he$ and $\li7$. The baryon to photon ratio which is
consistent with the observational data is restricted in the narrow
region, $\eta \simeq (4 - 6) \times 10^{-10}$. From
Fig.~\ref{fig:m100_eta_tr}(a), we find that $T_R \lesssim 3.7$~MeV is
excluded at 95 $\%$ C.L. for $B_h$= 1. On the other hand, from
Fig.~\ref{fig:m100_eta_tr}(b) we obtain the milder constraint that
$T_R \lesssim 2.5$~MeV is excluded at 95 $\%$ C.L. for $B_h =
10^{-2}$. In Fig.~\ref{fig:m0.01_eta_tr} we plot the contours of the
confidence level for $m_{\phi}= 10$ GeV in the same way as
Fig.~\ref{fig:m100_eta_tr}. Compared to Fig.~\ref{fig:m100_eta_tr}, as
we mentioned above, we find that the lower bound on the reheating
temperature becomes higher for a smaller mass.  From
Fig.~\ref{fig:m0.01_eta_tr} we get the lower bound on the reheating
temperature that $T_R \gtrsim 5.0$ MeV (4.0 MeV) at 95 $\%$ C.L. for
$B_h$= 1 ($B_h = 10^{-2}$)

In Fig.~\ref{fig:m_tr} the lower bound on $T_R$ as a function of
$m_{\phi}$ are plotted for (a) $B_h$ = 1 and (b) $B_h = 10^{-2}$. The
solid line denotes 95 $\%$ C.L. and the dotted line denotes 68 $\%$
C.L. As it is expected, the curve of the lower bound on $T_R$ is a
gentle monotonic decreasing function of $m_{\phi}$. In
Fig.~\ref{fig:m_tr}(a), we can see that $T_R$ should be higher than 4
MeV at 95 $\%$ C.L. for $B_h$ = 1 in $m_{\phi}$ = 10 GeV - $10^2$
TeV\footnote{Though we have adopted the experimental error of the each
hadron interaction cross section as 50$\%$ in the Monte Carlo
simulation because of no data, the lower bound on $T_R$ might become
about 10$\%$ higher than the above values if we adopt the more sever
experimental error as 10$\%$ instead of 50$\%$.}. On the other hand,
in Fig.~\ref{fig:m_tr}(b) we find that the constraint gets milder for
$B_h = 10^{-2}$. It is shown that $T_R \lesssim 2.5$ MeV is excluded
at 95 $\%$ C.L. for $B_h = 10^{-2}$. In Fig.~\ref{fig:tr_nnu} we find
that $N_{\nu}^{\rm eff}$ can be allowed as small as 2.8 for $B_h$ = 1
( 1.9 for $B_h = 10^{-2}$).

%%%%%%%%%%%%%%%%%%%%%%%%%%%%%%%%%%%%%%%%%%%%%%%%%%%%%%%%%%%%%%%%%%%%%%
\subsection{Summary of hadron injection}
%%%%%%%%%%%%%%%%%%%%%%%%%%%%%%%%%%%%%%%%%%%%%%%%%%%%%%%%%%%%%%%%%%%%%%

In this section we have seen that the BBN constraint on the reheating
temperature becomes much more stringent if a massive particle has a
branching to hadrons. For successful BBN the reheating temperature
should be higher than $2.5 - 4$~MeV for the branching ratio $B_h = 1
- 10^{-2}$. The hadron injection generally occurs if the late-time
reheating is caused by the heavy particle with mass larger than $\sim
1$~GeV. Many candidates for the late-time reheating such as SUSY
particles and flatons have such large masses and hence the constraint
obtained here is crucial in constructing particle physics models based
on SUSY or thermal inflation models.

For the lower limit of the reheating temperature, the effective number
of the neutrino species $N_{\nu}^{\rm eff}$ is given by 2.8 and 1.9
for $B_h$ = 1 and $10^{-2}$, respectively.  Since the limiting
temperature is close to the neutrino decoupling temperature, the
deviation of $N_{\nu}^{\rm eff}$ from the standard value (i.e. 3) is
small and hence the detection may not be easy.

However, from more general point of view, it is possible that light
particles with mass $\lesssim 1$~GeV are responsible for the late-time
reheating. In this case, as seen in the previous section, the
reheating temperatures as low as $\sim 0.7$~MeV are allowed. For such
low reheating temperature, neutrinos cannot be produced
sufficiently. Thus the effective number of the neutrino species
$N_{\nu}^{\rm eff}$ becomes much less than $3$. This leads to very
interesting effects on the formation of large scale structures and CMB
anisotropies, which we discuss in the next section.

%%%%%%%%%%%%%%%%%%%%%%%%%%%%%%%%%%%%%%%%%%%%%%%%%%%%%%%%%%%%%%%%%%%%%%
\section{Constraints from Large Scale Structure and CMB anisotropy}
%%%%%%%%%%%%%%%%%%%%%%%%%%%%%%%%%%%%%%%%%%%%%%%%%%%%%%%%%%%%%%%%%%%%%%
\label{sec:LSS_CMB}

In this section, we discuss possibility to set constraints on the
late-time entropy production from the large scale structure and CMB
anisotropies.  Hereafter, we only consider flat universe models with
cosmological constant which are suggested by recent distant Supernovae
(SNe) surveys~\cite{Riess,Perl} and measurements of CMB
anisotropies~\cite{CMB}.

The late-time entropy production influences formation of the large
scale structure and CMB anisotropies since the matter-radiation 
equality epoch is shifted if the effective number of neutrino species 
changes.  The ratio of neutrino density to black-body photon density 
is $\rho_\nu / \rho_\gamma = (7/8)(4/11)^{4/3} N_\nu$. 
Therefore the redshift of matter-radiation equality can be written as 
a function of $N_\nu$: 
\begin{equation}
\label{eq:equality}
1+z_{\rm eq} = 4.03 \times 10^{4} \Omega_0 h^2 
\left( 1+ {7 \over 8} \left({4 \over 11}\right)^{4/3}N_\nu \right)^{-1} ,
\end{equation}
where $\Omega_0$ is the density parameter and $h$ is the non-dimensional 
Hubble constant normalized by $100\rm km/s/Mpc$.

Let us now discuss distribution of galaxies on large scales.  
For a quantitative analysis, we define the matter power spectrum
in Fourier space as $P(k) \equiv <\vert \delta_k
\vert^2>$, where $\delta_k$ is the Fourier transform of density
fluctuations and $<>$ denotes the ensemble average.  Hereafter, we
assume the Harrison-Zel'dovich power spectrum, which is motivated by
the inflation scenario, as an initial shape of the power spectrum,
i.e., $P(k) \propto k$.  As fluctuations evolve in the expanding
universe, the shape of the power spectrum is changed.  
One often introduces the transfer function $T(k)$ to describe 
this modification of the initial power spectrum as 
$P(k) = A k T(k)^2$, where $A$ is an arbitrary constant.  
In case of
standard cold dark matter (CDM) dominated models, Bardeen et al.~\cite{Bard}
found a fitting formula:
\begin{equation}
T(k)={\ln (1+2.34q)\over 2.34q}[1+3.89q+(16.1q)^2 + 
(5.46q)^3 + (6.71q)^4 ]^{-1/4} ,
\end{equation}
where $q =k/\Omega_0 h^2 \rm Mpc^{-1}$ when the baryon density is
negligible small compared to the total density.  It is easy to explain
why $q$ is parameterized by $\Omega_0 h^2$.  This is because CDM
density fluctuations cannot evolve and stagnate during a radiation
dominated era. Only after the matter-radiation equality epoch,
fluctuations can evolve.  Therefore the CDM power spectrum has a peak
which corresponds to the horizon scale of the matter-radiation
equality epoch.  In fact, the wave number of the horizon scale at the
equality epoch can be written as $k_{\rm eq} = \sqrt{2\Omega_0
(1+z_{\rm eq})}H_0$, where $H_0$ is the Hubble constant at present,
that is proportional to $\Omega_0 h^2$.  In the actual observations,
distances in between galaxies are measured in the units of $h^{-1}\rm
Mpc$.  Therefore to fit the observational data by the CDM type power
spectrum, we usually introduce so-called {\it shape parameter}
$\Gamma_{\rm s} = \Omega_0 h$.  It is known that we can fit the galaxy
distribution if $\Gamma_{\rm s} \simeq 0.25\pm 0.05$~\cite{PD} which
suggests a low density universe.  If the late-time entropy production
takes place, however, we need to take into account $N_\nu^{\rm eff}$
dependence of the matter-radiation equality epoch
(Eq.~(\ref{eq:equality})).  Therefore $\Gamma_{\rm s}$ should be
written as
\begin{equation}
\Gamma_{\rm s} = 1.68 \Omega_0 h/(1+0.227N_\nu^{\rm eff}) .
\end{equation}
We plot contours of $\Gamma_{\rm s}$ on $\Omega_0-N_\nu^{\rm eff}$ plane
in Fig.~\ref{fig:gamma}.  It is shown that 
smaller $\Omega_0$ is preferable for $N_\nu^{\rm eff} < 3$ 
with the same value of $\Gamma_{\rm s}$.
We also plot the power spectra for $\Omega_0=0.3$ and $h=0.7$ with 
different $N_\nu^{\rm eff}$'s in Fig.~\ref{fig:power}.  Here 
we do not simply employ the fitting formula but 
numerically solve the evolution 
of density fluctuations~\cite{SG}. 
It is shown that the peak location of 
a model with smaller $N_\nu^{\rm eff}$ shifts to the smaller scale 
(larger in $k$) since 
smaller $N_\nu^{\rm eff}$ makes 
the equality epoch earlier which means the horizon scale at the equality
epoch becomes smaller.  
We have hope that current larege scale structrue surveys such as 2DF and 
Sloan Digital Sky Survey (SDSS) may determine the precise value of 
$\Gamma_{\rm s}$.

%Therefore the constraint from the large scale galaxy distribution becomes 
%much tighter with $N_\nu^{\rm eff} < 3$  (see Fig. ).   

Besides the shape of the power spectrum, the amplitude is another
important observational quantities to test models.  
On very large scales, the amplitude of the power spectrum is 
determined by CMB anisotropies which are measured by COBE/DMR~\cite{COBE}. 
Since COBE/DMR scales are much larger than the horizon scale 
of the matter-radiation equality epoch, however, 
it is not sensitive to the transfer function $T(k)$ but the 
overall amplitude $A$. 
In order to  compare the expected 
amplitude of the power spectrum from each CDM model with 
large scale structure observations, 
we employ the specific mass
fluctuations within a sphere of a radius of $8h^{-1}\rm Mpc$, i.e.,
$\sigma_8$ which is defined as 
\begin{eqnarray}
\sigma_8^2 & = & <(\delta M / M(R))^2 >_{R = 8h^{-1}{\rm Mpc}}  \nonumber\\
& = & {1\over 2\pi^2} \int dk k^2 P(k) W(kR)^2|_{R= 8h^{-1}{\rm Mpc}}  ,
\end{eqnarray}
where $W(kR)$ is a window function for which we employ a top hat shape
as $W(kR) \equiv 3\left( \sin(kR)-kR\cos(kR)\right)/(kR)^3$.  Eke et
al.~\cite{Eke} obtained the observational value of $\sigma_8$ which is deduced
from the rich cluster abundance at present as 
\begin{equation}
\label{eq:sigma8obs}
\sigma_8 = (0.52\pm
0.04)\Omega_0^{-0.52+0.13\Omega_0} .  
\end{equation}
Other estimates of 
$\sigma_8$~\cite{sigma8} are agreed with their result.  
For CDM models with standard thermal history, the value of $\sigma_8$ 
is a function of $\Omega_0$ and $h$.  With the late-time reheating,
however, $\sigma_8$ for fixed $\Omega_0$ and $h$ 
becomes larger. The reason is following.   
Since we fix $\Omega_0$ and $h$, the normalization 
factor $A$ is same regardless of the value of $N_\nu^{\rm eff}$.
As is shown in Fig.~\ref{fig:power}, the amplitude of the power spectrum 
on $8h^{-1}\rm Mpc$, i.e., $\sigma_8$, is larger for smaller 
$N_\nu^{\rm eff}$.  
In Fig.~\ref{fig:sigma8}, we show the allowed region on the $\Omega_0-h$ 
plane for $N_\nu^{\rm eff} = 0.5, 2$ and $3$ for COBE-normalized 
flat CDM models with the Harrison-Zel'dovich spectrum. 
The shaded region satisfies the matching 
condition with the cluster abundance Eq.~(\ref{eq:sigma8obs}).  
For fixed $h$, models with smaller $N_\nu^{\rm eff}$
prefer lower $\Omega_0$.  Recently, the HST key project 
on the Extragalactic Distance Scale has reported that 
$h=0.71\pm 0.06$ (1$\sigma$) by using various 
distant indicators~\cite{Mould}.
  From SNe measurements, $\Omega_0 = 0.28\pm 0.8$ 
for flat models (see Fig.~7 of \cite{Perl}).
CDM models with $N_\nu^{\rm eff} = 0.5 \sim 3$ are 
still consistent with above value of $h$ and $\Omega_0$.  
However we  expect further precise determination of $\Omega_0$, $h$ (from 
distant SNe surveys and measurements of CMB anisotropies) and 
$\sigma_8$ (from 2DF or SDSS)
will set a stringent constraint on $N_\nu^{\rm eff}$ and $T_R$ in near future.

Finally we discuss the CMB constraint on $T_R$. 
%We expand the CMB temperature anisotropies $\Delta T/T$ into 
%multipole components as $\Delta T/T = \sum_\ell a_\ell P_ell(\mu)$.  
Let us introduce temperature angular power spectrum $C_\ell$ where 
$\ell$ is the multipole number of the spherical harmonic 
decomposition.  The rms temperature anisotropy 
of CMB can be written as 
$<\vert\Delta T/T\vert^2> = \sum_\ell(2\ell+1)C_\ell /4\pi$.
Using $C_\ell$, we can extract various important information 
of cosmology, such as the curvature of the universe, $\Omega_0$,
cosmological constant, $h$ and so on (see, e.g., \cite{Hu}).
In fact, we can measure the matter radiation equality epoch by 
using the height of peaks of $C_\ell$.  The peaks are boosted during 
the matter-radiation equality epoch.  If the matter-radiation equality 
is earlier, the correspondent horizon scale is smaller.  
Therefore we expect lower heights for first one or two peaks 
since these peaks are larger than the horizon scale at the 
equality epoch and do not suffer the boost as is shown in 
Fig.~\ref{fig:CMB}.
With the present angular
resolutions and sensitivities of COBE observation~\cite{COBE} or current
balloon and ground base experiments, however, it is impossible to set a
constraint on $N_{\nu}^{\rm eff}$. 
It is expected that future satellite
experiments such as MAP~\cite{MAP} and PLANCK~\cite{PLANCK} will give us
a useful information about $N_{\nu}^{\rm eff}$. From Lopez et al.'s
analysis~\cite{Lopez}, MAP and PLANCK have sensitivities that $\delta
N_{\nu}^{\rm eff} \gtrsim 0.1$ (MAP) and $0.03$ (PLANCK) 
including polarization
data, even if all cosmological parameters are determined
simultaneously (see also Fig.~\ref{fig:CMB}). 
   From such future observations of anisotropies of CMB,
it is expected that we can precisely determine $T_R$.
%%

%%%%%%%%%%%%%%%%%%%%%%%%%%%%%%%%%%%%%%%%%%%%%%%%%%%%%%%%%%%%%%%%%%%%%%
\section{Conclusion}
%%%%%%%%%%%%%%%%%%%%%%%%%%%%%%%%%%%%%%%%%%%%%%%%%%%%%%%%%%%%%%%%%%%%%%
\label{sec:conclude}
In this paper we have investigated the various cosmological effects
induced by the late-time entropy production due to the massive
particle decay. The neutrino distribution functions have been obtained
by solving the Boltzmann equations numerically. We have found that if
the large entropy are produced at about $t \simeq 1$ sec, the
neutrinos are not thermalized very well and hence do not have the
perfect Fermi-Dirac distribution. The deficits of the neutrino
distribution functions due to the insufficient thermalization decrease
the Hubble expansion rate and weakens the weak interaction rates
between proton and neutron.  The above two effects changes the
freeze-out value of $n/p$ significantly.  Especially the produced
$\4he$ mass fraction $Y$ is so sensitive to $n/p$ that the predicted
value of $Y$ is changed drastically.  Comparing the theoretical
predictions of D, $\4he$ and $\li7$ to the observational data, we have
estimated the lower bound on the reheating temperature $T_R$ after the
entropy production.  We have found that $T_R \lesssim 0.7$~MeV is
excluded at 95 $\%$ C.L. In other wards, $T_R$ can be as low as 0.7
MeV. Then the effective number of neutrino species $N_{\nu}^{\rm eff}$
can be as small as $0.1$. It is enough sensitive for 
the ongoing large scale structure observations such as 2DF and SDSS or future  
satellite experiments (MAP and PLANCK) of CMB anisotropies 
to detect such modifications on
$N_{\nu}^{\rm eff}$ and we can find out the vestige of the late-time
entropy production.
 
Furthermore, we have also studied the case in which the massive
particle has some decay modes into quarks or gluons. In this scenario,
a lot of hadrons, {\it e.g.}  pions, kaons, protons and neutrons,
which are originated by the fragmentation of the high energy quarks
and gluons are injected into thermal bath. The emitted hadrons
extraordinarily inter-convert the ambient protons and neutrons each
other through the strong interaction even after the freeze-out time of
the neutron to proton ratio $n/p$. Then the predicted value of $Y$
increase extremely and we can constrain $T_R$ and the branching ratio
of the hadronic decay mode $B_h$ comparing to the observational light
element abundances. We have found $T_R$ should be higher than 2.5 MeV
- 4 MeV at 95 $\%$ C.L. for $B_h$ = $10^{-2}$ - 1. The above results
tells us that $N_{\nu}^{\rm eff}$ can be as small as 1.9 - 2.8 even in
the hadron injection scenario for $B_h = 10^{-2}$ - 1. Then it still
may be possible to detect the modifications on $N_{\nu}^{\rm eff}$ by
MAP and PLANCK.

%%%%%%%%%%%%%%%%%%%%%%%%%%%%%%%%%%%%%%%%%%%%%%%%%%%%%%%%%%%%%%%%%%%%%%
\section*{Acknowledgment}
%%%%%%%%%%%%%%%%%%%%%%%%%%%%%%%%%%%%%%%%%%%%%%%%%%%%%%%%%%%%%%%%%%%%%%
K.K. wish to thank J. Yokoyama and T. Asaka for useful discussions.
This work was partially supported by the Japanese Grant-in-Aid for
Scientific Research from the Monbusho, Nos.\ 10640250~(MK), 10-04502
(KK), 9440106~(NS) and ``Priority Area: Supersymmetry and Unified
Theory of Elementary Particles(\#707)''(MK) and by the Sumitomo
Foundation (NS).

%%%%%%%%%%%%%%%%%%%%%%%%%%%%%%%%%%%%%%%%%%%%%%%%%%%%%%%%%%%%%%%%%%%%%%
\appendix
%%%%%%%%%%%%%%%%%%%%%%%%%%%%%%%%%%%%%%%%%%%%%%%%%%%%%%%%%%%%%%%%%%%%%%
\label{sec:appendix}
\section*{Reduction of collision integral}

This appendix shows how we can reduce the nine dimensional integrals
in Eq.~(\ref{eq:collision}) of the collision term $C_{i, \rm scat}$
for the scattering process into one dimensional integrals. Notice
that, since we treat the massless neutrino, the norm of the neutrino
momentum equals to its energy $|\mbox{\boldmath $p_i$}| = E_i$.  Here
we divide the collision term into two parts:
\begin{equation}
     \label{dev_coll}
     C_{i, \rm scat} = - F  + B,
\end{equation}
where $F$ represents the forward process and $B$ represents the
backward process. They are given by
\begin{equation}
     \label{forward_rate}
     F = \frac{g_e}{2E_1}\int \frac{dp_2^3}{2E_2(2\pi)^3}\int
     \frac{dp_3^3}{2E_3(2\pi)^3} \int \frac{dp_4^3}{2E_4(2\pi)^3}
     (2\pi)^4 \delta^4(p_1+p_2-p_3-p_4) S|M|^2 \Lambda_F,
\end{equation}
\begin{equation}
     \label{eq:backward_rate}
     B = \frac{g_e}{2E_1}\int \frac{dp_2^3}{2E_2(2\pi)^3}\int
     \frac{dp_3^3}{2E_3(2\pi)^3} \int \frac{dp_4^3}{2E_4(2\pi)^3}
     (2\pi)^4 \delta^4(p_1+p_2-p_3-p_4) S|M|^2 \Lambda_B,
\end{equation}
where $g_e$ = 2 and the phase space factors are given by
\begin{eqnarray}
     \label{eq:Lambda_fb}
     \Lambda_F &=&
     f_1(E_1)f_2(E_2)\left(1-f_3(E_3)\right)\left(1-f_4(E_4)\right), \\
     \Lambda_B &=&
     \left(1-f_1(E_1)\right)\left(1-f_2(E_2)\right)f_1(E_3)f_2(E_4).
\end{eqnarray}

The integral over $d^3p_4$ is immediately done using
$\delta^3(\mbox{\boldmath $p_1 + p_2 - p_3- p_4$})$. From the momentum
conservation, $\mbox{\boldmath $|p_4|$}$ is given by
\begin{equation}
     \label{eq:p4}
     |\mbox{\boldmath $p_4$}|^2 = E_4^2 = E_2^2+2E_2R\cos{\eta}+R^2,
\end{equation}
where $\mbox{\boldmath $R$}$ $\equiv$ $\mbox{\boldmath $p_1 - p_3$}$, $R$
= $|\mbox{\boldmath $R$}|$ and $\cos{\eta}$ $\equiv$ $\mbox{\boldmath $R
\cdot p_2$}$/($|\mbox{\boldmath $p_2$}|$R).

The remaining delta function $\delta(E_1+E_2-E_3-E_4)$ shows the
energy conservation low which is given by
\begin{equation}
     \label{eq:energy_cons}
     E_4^2 = E_1^2 + E_2^2 + E_3^2 + 2(E_1E_2-E_1E_3-E_2E_3).
\end{equation}
We can generally take the momentum axes as
\begin{eqnarray}
     \label{eq:axes}
     \mbox{\boldmath $R$} &=& (0, 0, R) \\
     \mbox{\boldmath $p_2$} &=& (E_2\sin{\eta}\sin{\phi},
     E_2\sin{\eta}\cos{\phi}, E_2\cos{\eta}), \\
     \mbox{\boldmath $p_3$} &=& (E_3\sin{\xi}, 0, E_3\sin{\xi}),
\end{eqnarray}
where
\begin{equation}
     \label{eq:cosxi}
     \cos{\xi}= \frac{E_1^2 - E_3^2 - R^2}{2E_3R}.
\end{equation}
Then $|\cos{\xi}| \le 1$ demands
\begin{equation}
     \label{eq:int_xi}
     |E_1-E_3| \le R \le E_1+E_3.
\end{equation}
The volume element of $\mbox{\boldmath $p_2$}$ is given by $dp_2^3 =
E_2^2d\cos{\eta}d{\phi}$ and from Eqs.~(\ref{eq:p4}) and
(~\ref{eq:energy_cons}) the azimuthal angle is obtained by
\begin{eqnarray}
     \label{eq:angles}
     \cos{\eta}=- \frac{R^2-(E_1-E_3)^2-2E_2(E_1-E_3)}{2E_2R}.
\end{eqnarray}
Then $|\cos{\eta}|\le 1$ demands
\begin{equation}
     \label{eq:int_eta}
     |E_1-E_3| \le R \le E_1 + 2 E_2 - E_3.
\end{equation}
 From Eqs.~(\ref{eq:int_xi}) and ~(\ref{eq:int_eta}), we can obtain the
allowed region of R,
\begin{equation}
     \label{eq:R_int}
     |E_1-E_3| \le R \le {\rm Inf}[E_1+E_3, E_1+2E_2-E_3].
\end{equation}
Since the volume element of $\mbox{\boldmath $p_3$}$ is given by
$dp_3^2 = 2\pi E_3^2dE_3d\cos{\theta}$ where
$\cos{\theta}=\mbox{\boldmath{$p_1\cdot p_3$}}/(E_3E_1)$, the
differential angle element is evaluated by
\begin{equation}
     \label{eq:dR}
     d\cos{\theta}= - \frac{R}{2 E_1E_3}dR.
\end{equation}
 From Eq.~(\ref{eq:R_int}) we can see that the integration can be performed
in the four allowed intervals,
\begin{eqnarray}
     \label{eq:four_int}
     - F  + B &=& \frac1{128E_1^2}\int_0^{\infty}dE_3
     \int_0^{\infty}dE_2 \int dR \int_0^{2\pi}\frac{d\phi}{2\pi} |M|^2
     (-\Lambda_F + \Lambda_B) \nonumber \\
              &=& \frac1{128E_1^2}
     \left[\int_{0}^{E_1}dE_3\int_{0}^{E_3}dE_2\int_{E_1-E_3}^{E_1+2E_2-E_3}dR
     +\int_{0}^{E_1}dE_3\int_{E_3}^{\infty}dE_2\int_{E_1-E_3}^{E_1+E_3}dR
     \nonumber \right.\\
     & &\left. \hspace{1cm} +\int_{E_1}^{\infty}dE_3\int_{-E_1+E_3}^{E_3}dE_2
                             \int_{-E_1+E_3}^{E_1+2E_2-E_3 }dR
     +\int_{E_1}^{\infty}dE_3\int_{E_3}^{\infty}dE_2
                              \int_{-E_1+E_3}^{E_1+E_3}dR\right]
      \nonumber \\
     & &\hspace{1cm} \times \int_0^{2\pi}\frac{d\phi}{2\pi}S|M|^2
     (-\Lambda_F + \Lambda_B).
\end{eqnarray}
Even though we only show the case of of $\nu_e$ here, we can get the
same procedure for $\nu_{\mu}$ and $\nu_{\tau}$ if $C_V$ and $C_A$ are
replaced by $\tilde{C_V}$ and $\tilde{C_A}$. As we also noted in
Sec~\ref{sec:formulation}, we assume that electrons obey the
Boltzmann distribution function $e^{-E/T}$.  In addition, since
neutrinos are massless, the energy momentum conservation gives $p_1
\cdot p_4=p_2 \cdot p_3$ in the elastic scattering process. The above
assumptions simplify the integrations still more.

For the forward reaction, $\nu(p_1)+e^{\pm}(p_2) \rightarrow
\nu(p_3)+e^{\pm}(p_4)$, the phase space factor is given by
\begin{equation}
     \label{red_lambda_f}
     \Lambda_F = f_{\nu}(E_1)\left(1-f_{\nu}(E_3)\right)\exp{[-\frac{E_2}{T}]}.
\end{equation}
Then $F_1$ and $F_2$ in Eq.~(\ref{eq:C-scat}) are analytically estimated as
\begin{eqnarray}
     \label{f_1_def}
     F_1 &\equiv&
          \left[\int_{0}^{E_3}dE_2\int_{E_1-E_3}^{E_1+2E_2-E_3}dR +
                \int_{E_3}^{\infty}dE_2\int_{E_1-E_3}^{E_1+E_3}dR
              \right]
          \int_0^{2\pi}\frac{d\phi}{2\pi} \cdot
            \frac{S|M|^2 e^{-\frac{E_2}{T}}}{256(C_V^2+C_A^2)G_F^2}
          \nonumber \\
         &=& 2T^4 \left[ E_1^2+E_3^2+2T(E_1-E_3)+4T^4 \right]
            -T^2\left[  E_1^2E_3^2+2E_1E_3(E_1+E_3)T \right.
            \nonumber \\
         & &  \hspace{4cm} \left. +2(E_1+E_3)^2T^2 +
           4(E_1+E_3)T^3 + 8T^4  \right] e^{-\frac{E_3}{T}},
\end{eqnarray}
\begin{eqnarray}
     \label{f_2_def}
     F_2 &\equiv&
     \left[\int_{-E_1+E_3}^{E_3}dE_2\int_{-E_1+E_3}^{E_1+2E_2-E_3}dR +
                \int_{E_3}^{\infty}dE_2\int_{-E_1+E_3}^{E_1+E_3}dR
              \right]
          \int_0^{2\pi}\frac{d\phi}{2\pi}\cdot
            \frac{S|M|^2 e^{-\frac{E_2}{T}}}{256(C_V^2+C_A^2)G_F^2}
          \nonumber \\
         &=&
         2T^4(E_1^2+E_3^2-2T(E_1-E_3)+4T^4) e^{\frac{E_1-E_3}{T}}
         -T^2 \left[ E_1^2E_3^2+2E_1E_3(E_1+E_3)T  \right.
            \nonumber \\
         & &  \left. \hspace{4cm} +2(E_1+E_3)^2T^2 +
           4(E_1+E_3)T^3 + 8T^4 \right] e^{-\frac{E_3}{T}}.
\end{eqnarray}

On the other hand, for the backward reaction, $\nu(p_1)+e^{\pm}(p_2)
\leftarrow \nu(p_3)+e^{\pm}(p_4)$, the phase space factor is given by,
\begin{equation}
     \label{red_lambda_B}
     \Lambda_B =
     \left(1-f_{\nu}(E_1)\right)f_{\nu}(E_3)\exp{(-\frac{E_1+E_2+E_3}{T})}.
\end{equation}
Then we can analytically obtain $B_1$ and $B_2$ in Eq.~(\ref{eq:C-scat})
as
\begin{eqnarray}
     \label{b_1_def}
     B_1 &\equiv& \left[\int_{0}^{E_3}dE_2\int_{E_1-E_3}^{E_1+2E_2-E_3}dR +
                  \int_{E_3}^{\infty}dE_2\int_{E_1-E_3}^{E_1+E_3}dR\right]
                 \int_0^{2\pi}\frac{d\phi}{2\pi}\cdot
            \frac{S|M|^2 e^{-\frac{E_1+E_2+E_3}{T}}}{256(C_V^2+C_A^2)G_F^2},
            \nonumber \\
           &=&
           2T^4(E_1^2+E_3^2+2T(E_1-E_3)+4T^4)e^{-\frac{E_1-E_3}{T}}
         -T^2
         \left[E_1^2E_3^2+2E_1E_3(E_1+E_3)T  \right.
            \nonumber \\
         & &  \hspace{4cm} \left. +2(E_1+E_3)^2T^2 +
           4(E_1+E_3)T^3 + 8T^4 \right] e^{-\frac{E_1}{T}},
\end{eqnarray}
\begin{eqnarray}
     \label{b_2_def}
     B_2 &\equiv&
          \left[\int_{-E_1+E_3}^{E_3}dE_2\int_{-E_1+E_3}^{E_1+2E_2-E_3}dR +
                \int_{E_3}^{\infty}dE_2\int_{-E_1+E_3}^{E_1+E_3}dR
              \right]
          \int_0^{2\pi}\frac{d\phi}{2\pi} \cdot
            \frac{S|M|^2 e^{-\frac{E_1+E_2+E_3}{T}}}{256(C_V^2+C_A^2)G_F^2}
          \nonumber \\
           &=&
           2T^4\left( E_1^2+E_3^2-2T(E_1-E_3)+4T^4 \right) -T^2
          \left[ E_1^2E_3^2+2E_1E_3(E_1+E_3)T \right.
            \nonumber \\
         & &  \hspace{4cm} \left. +2(E_1+E_3)^2T^2 +
           4(E_1+E_3)T^3 + 8T^4 \right] e^{-\frac{E_1}{T}}.
\end{eqnarray}
%%

%%%%%%%%%%%%%%%%%%%%%%%%%%%%%%%%%%%%%%%%%%%%%%%%%%%%%%%%%%%%%%%%%%%%%%

%%%%%%%%%%%%%%%%%%%%%%%%%%%%%%%%%%%%%%%%%%%%%%%%%%%%%%%%%%%%%%%%%%%%%%
%TABLES
%%%%%%%%%%%%%%%%%%%%%%%%%%%%%%%%%%%%%%%%%%%%%%%%%%%%%%%%%%%%%%%%%%%%%%

\begin{table}[htbp]
     \begin{center}
         \leavevmode
         \begin{tabular}{ccccc|cc}
             && Process && & S$\left|M\right|^2$& \\ \hline
             & $\nu _{e} + e^{-} $
             &  $\rightarrow$
             & $ \nu _{e} + e^{-}$
             && $32G_F^2\left[(C_V + C_A)^2 (p_1 \cdot p_2)^2 + (C_V -
             C_A)^2 (p_1 \cdot 
               p_4)^2\right]$
             &\\
             & $\nu _{e} + e^{+} $
             &  $\rightarrow$
             & $ \nu _{e} + e^{+}$
             && $32G_F^2\left[(C_V - C_A)^2 (p_1 \cdot p_2)^2 + (C_V +
             C_A)^2 (p_1 \cdot 
               p_4)^2\right]$
             &\\            & $\nu _{e} + \bar{\nu}_{e} $
             &  $\rightarrow$
             &   $  e^{+} + e^{-}$
             && $32G_F^2\left[(C_V+C_A)^2 (p_1 \cdot p_4)^2 +
               (C_V-C_A)^2 (p_1 \cdot p_3)^2\right]$
             &\\
         \end{tabular}
         \caption{Matrix elements for electron neutrino
         interactions. $G_F$ is the Fermi coupling constant. Here we take
         $C_V=\frac12 + 2\sin^2\theta_W$, $C_A= \frac12 $ and the weak
         mixing angle $\sin^2\theta_W \simeq 0.231.$}
         \label{table:Mnue}
     \end{center}
\end{table}

\begin{table}[htbp]
     \begin{center}
         \leavevmode
         \begin{tabular}{ccccc|cc}
             && Process && & S$\left|M\right|^2$& \\ \hline
             & $\nu _{\mu} + e^{-} $
             &  $\rightarrow$
             & $ \nu _{\mu} + e^{-}$
             && $32G_F^2\left[(\tilde{C_V}+\tilde{C_A})^2 (p_1 \cdot
               p_2)^2 + (\tilde{C_V}-\tilde{C_A})^2 (p_1 \cdot
               p_4)^2\right]$
             &\\
             & $\nu _{\mu} + e^{+} $
             &  $\rightarrow$
             & $ \nu _{\mu} + e^{+}$
             && $32G_F^2\left[(\tilde{C_V}-\tilde{C_A})^2 (p_1 \cdot
               p_2)^2 + (\tilde{C_V}+\tilde{C_A})^2 (p_1 \cdot
               p_4)^2\right]$
             &\\            & $\nu _{\mu} + \bar{\nu}_{\mu} $
             &  $\rightarrow$
             &   $  e^{+} + e^{-}$
             && $32G_F^2\left[(\tilde{C_V}+\tilde{C_A})^2 (p_1 \cdot p_4)^2 +
               (\tilde{C_V}-\tilde{C_A})^2 (p_1 \cdot p_3)^2\right]$
             &\\
         \end{tabular}
         \caption{Matrix elements for muon neutrino or tau neutrino
         interactions. $G_F$ is the Fermi coupling constant.
         Here we take $\tilde{C_V}=C_V-1 = - \frac12 +
         2\sin^2\theta_W$, $\tilde{C_A}=C_A-1 = - \frac12 $ and the weak
         mixing angle $\sin^2\theta_W \simeq 0.231.$}
         \label{table:Mnumu}
     \end{center}
\end{table}

%%%%%%%%%%%%%%%%%%%%%%%%%%%%%%%%%%%%%%%%%%%%%%%%%%%%%%%%%%%%%%%%%%%%%%
%FIGURES
%%%%%%%%%%%%%%%%%%%%%%%%%%%%%%%%%%%%%%%%%%%%%%%%%%%%%%%%%%%%%%%%%%%%%%
%%
\begin{figure}
   \begin{center}
     \centerline{\psfig{figure=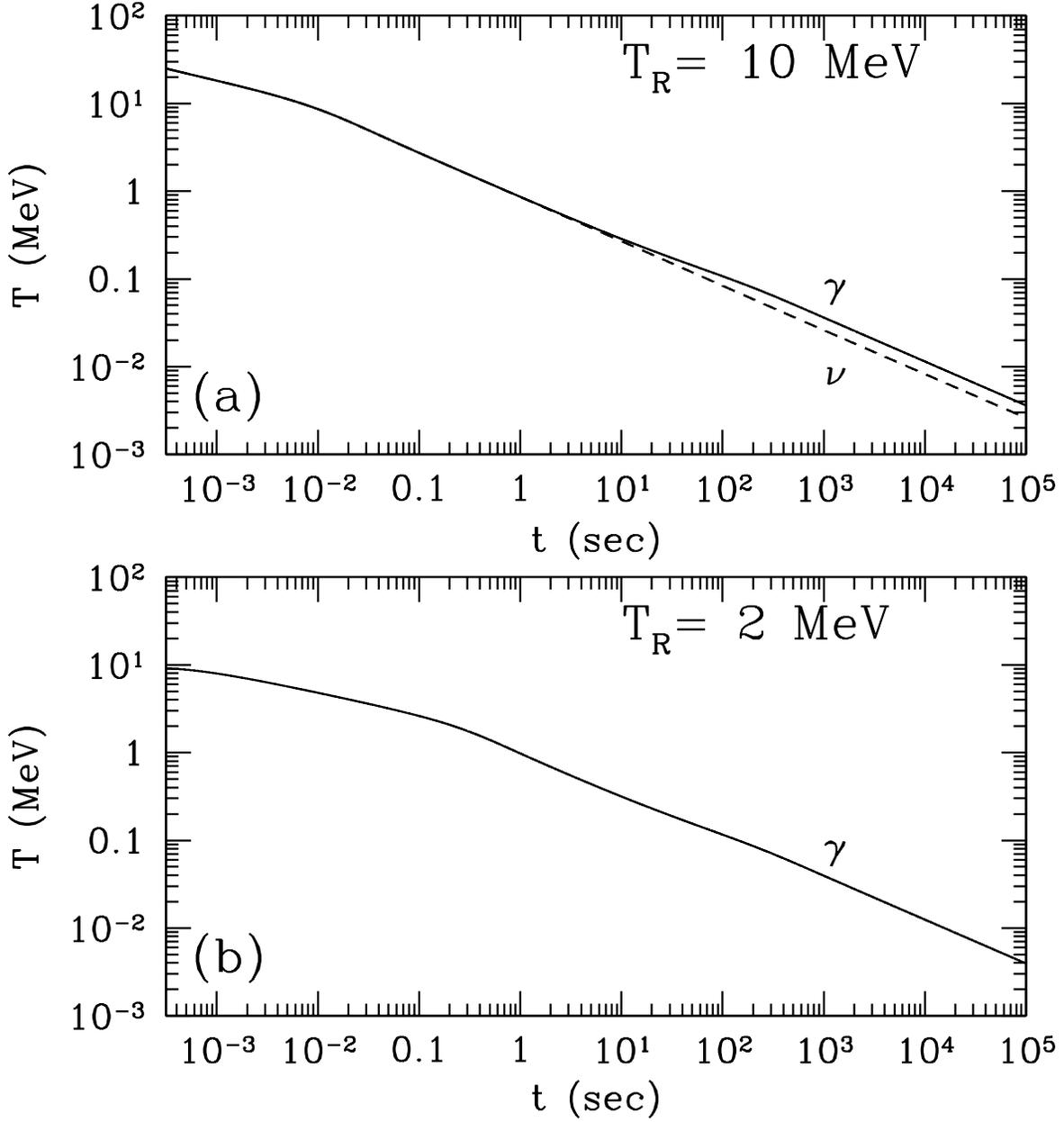,width=17cm}}
       \caption{%%
       Time evolution of the cosmic temperature (a) for $T_{R}=10$~MeV,
       and (b) for $T_{R}=2$~MeV. The dashed line denotes the neutrino
       temperature which can be defined only when they are thermalized
       sufficiently and have the perfect Fermi-Dirac distribution.}
       \label{fig:temp}
   \end{center}
\end{figure}
\newpage
\begin{figure}
   \begin{center}
     \centerline{\psfig{figure=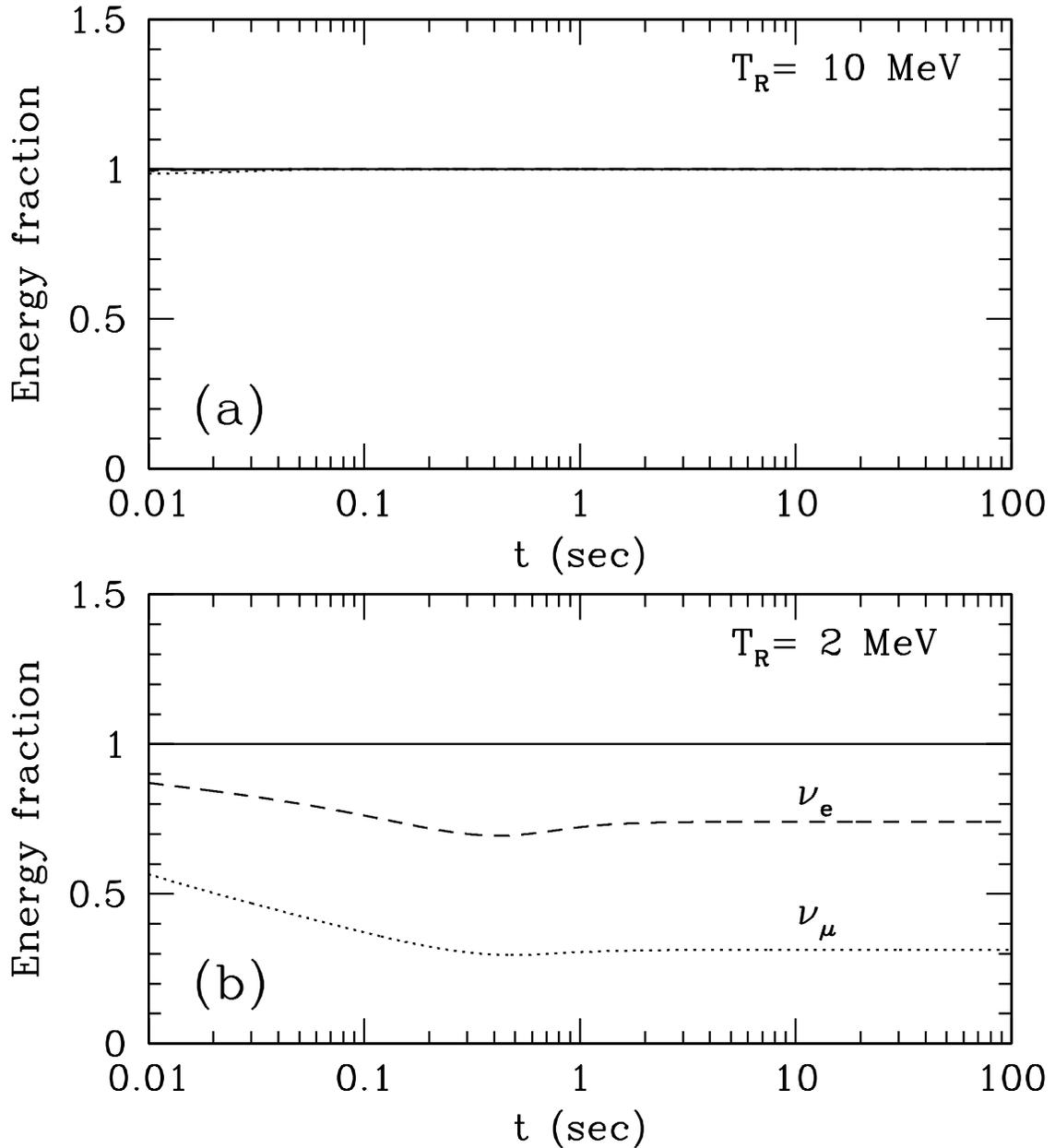,width=17cm}}
       \caption{%%
       Time evolution of the fraction of the energy density of
       $\nu_{e}$ (solid curve) and $\nu_{\mu}$ (dashed curve) to that
       of the standard big bang scenario for (a) $T_{R}= 10$~MeV and
       (b)$T_{R}=2$~MeV. Since the interaction of $\nu_{\tau}$ is as
       same as $\nu_{\mu}$, the curve of $\nu_{\mu}$ also represents
       $\nu_{\tau}$.}
       \label{fig:rho-nu}
   \end{center}
\end{figure}
\newpage
\begin{figure}
   \begin{center}
     \centerline{\psfig{figure=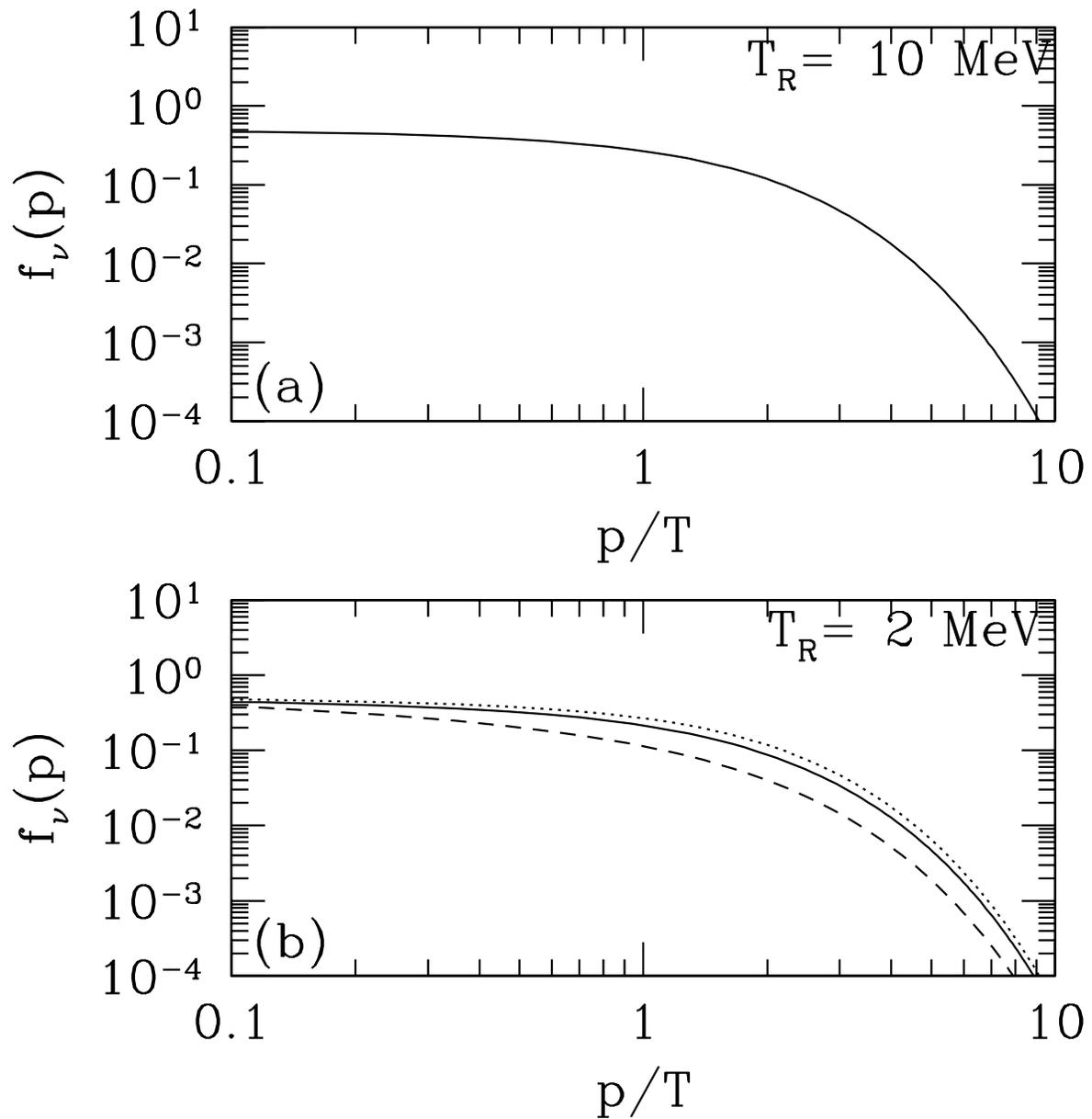,width=17cm}}
       \caption{%%
       Distribution function of $\nu_{e}$ (solid curve) and $\nu_{\mu}$
       (dashed curve) (a)for $T_{R}=10$~MeV and (b)for $T_{R}=2$~MeV.
       The dotted curve is the Fermi-Dirac distribution function.
       Since the interaction of $\nu_{\tau}$ is as same as $\nu_{\mu}$,
       the curve of $\nu_{\mu}$ also represents $\nu_{\tau}$.}
       \label{fig:distribution}
   \end{center}
\end{figure}
\newpage
\begin{figure}[htbp]
   \begin{center}
     \centerline{\psfig{figure=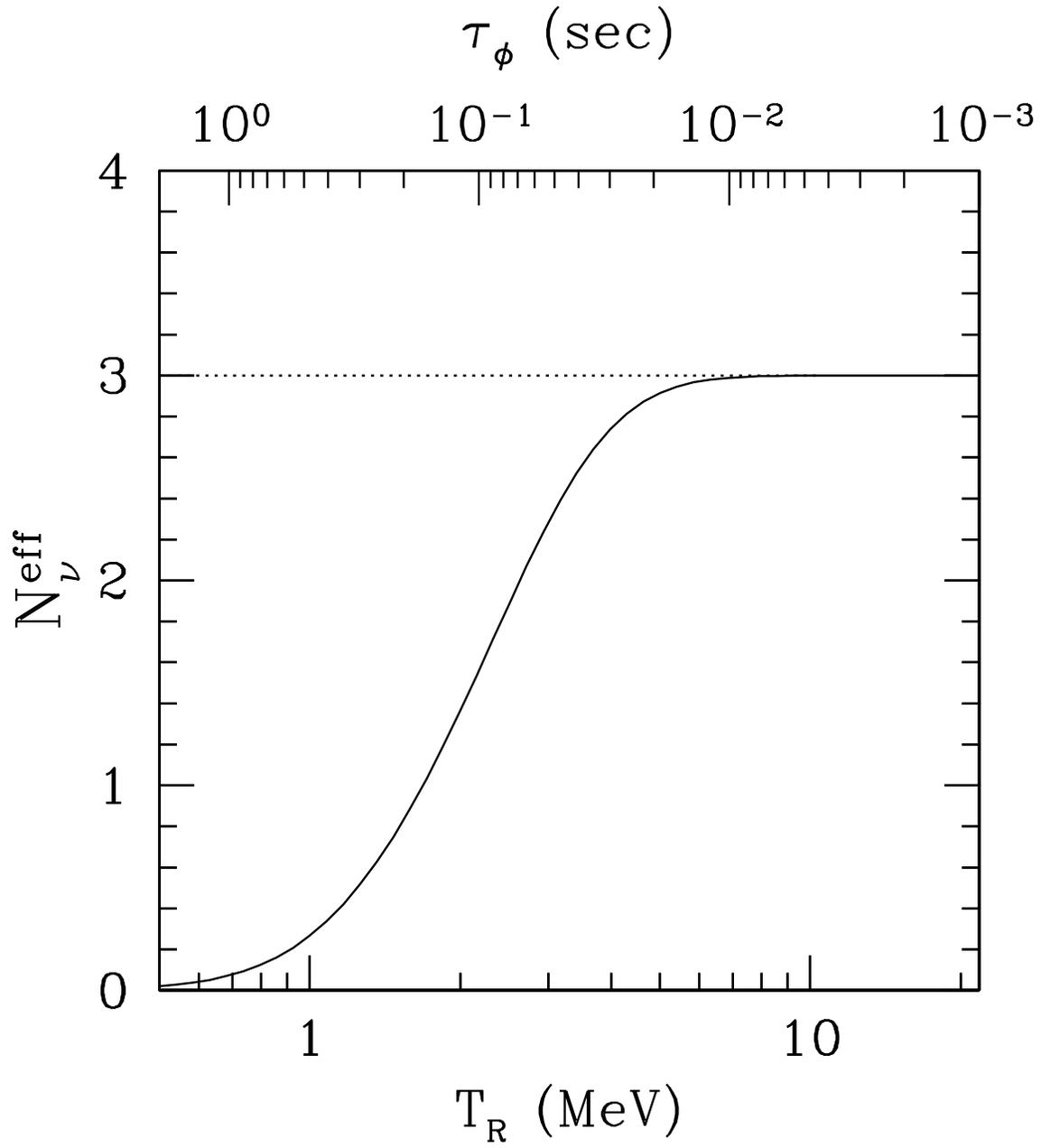,width=18.0cm}}
       \caption{%%
       Effective number of neutrino species $N_{\nu}^{\rm eff}$ as a
       function of reheating temperature $T_{R}$. The top horizontal
       axis denotes the lifetime which corresponds to $T_{R}$.}
       \label{fig:tr_nnu}
   \end{center}
\end{figure}
\newpage
\begin{figure}[htbp]
   \begin{center}
     \centerline{\psfig{figure=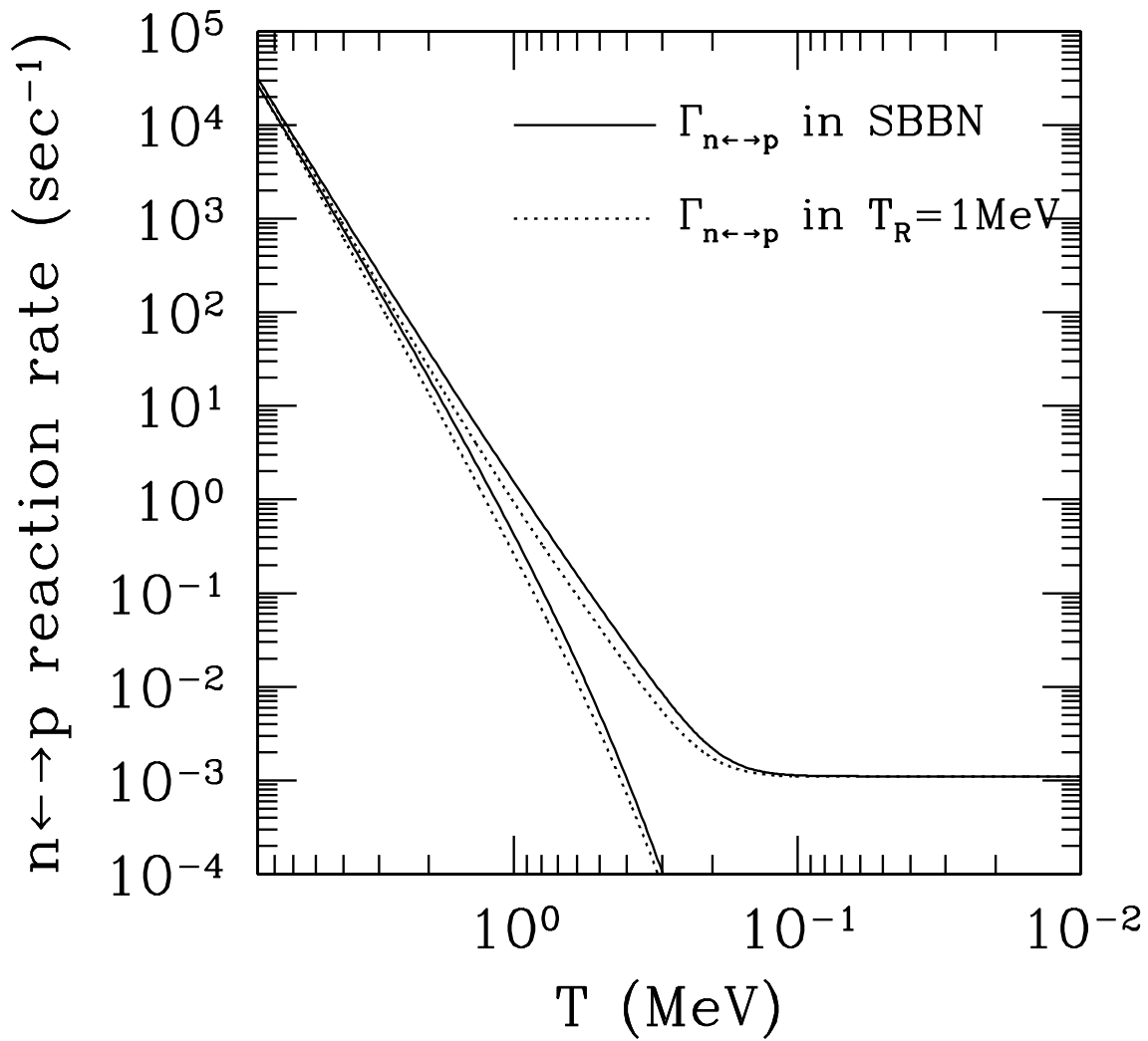,width=18.0cm}}
       \caption{%%
       Weak interaction rates (sec$^{-1}$) between neutron and proton.
       The upper curves are $\Gamma_{n \rightarrow p}$. The lower
       curves are $\Gamma_{p \rightarrow n}$.  The solid lines denote
       the case of $T_R = 10$ MeV which corresponds to the standard big
       bang scenario. The dotted lines denote the case of $T_R = 1$ MeV
       in the late-time entropy production scenario. Notice that $\Gamma_{n
       \rightarrow p}^{-1}$ reaches $\tau_n = 887$ sec in the low
       temperature.}
       \label{fig:weak_rate}
   \end{center}
\end{figure}
\newpage
\begin{figure}[htbp]
   \begin{center}
     \centerline{\psfig{figure=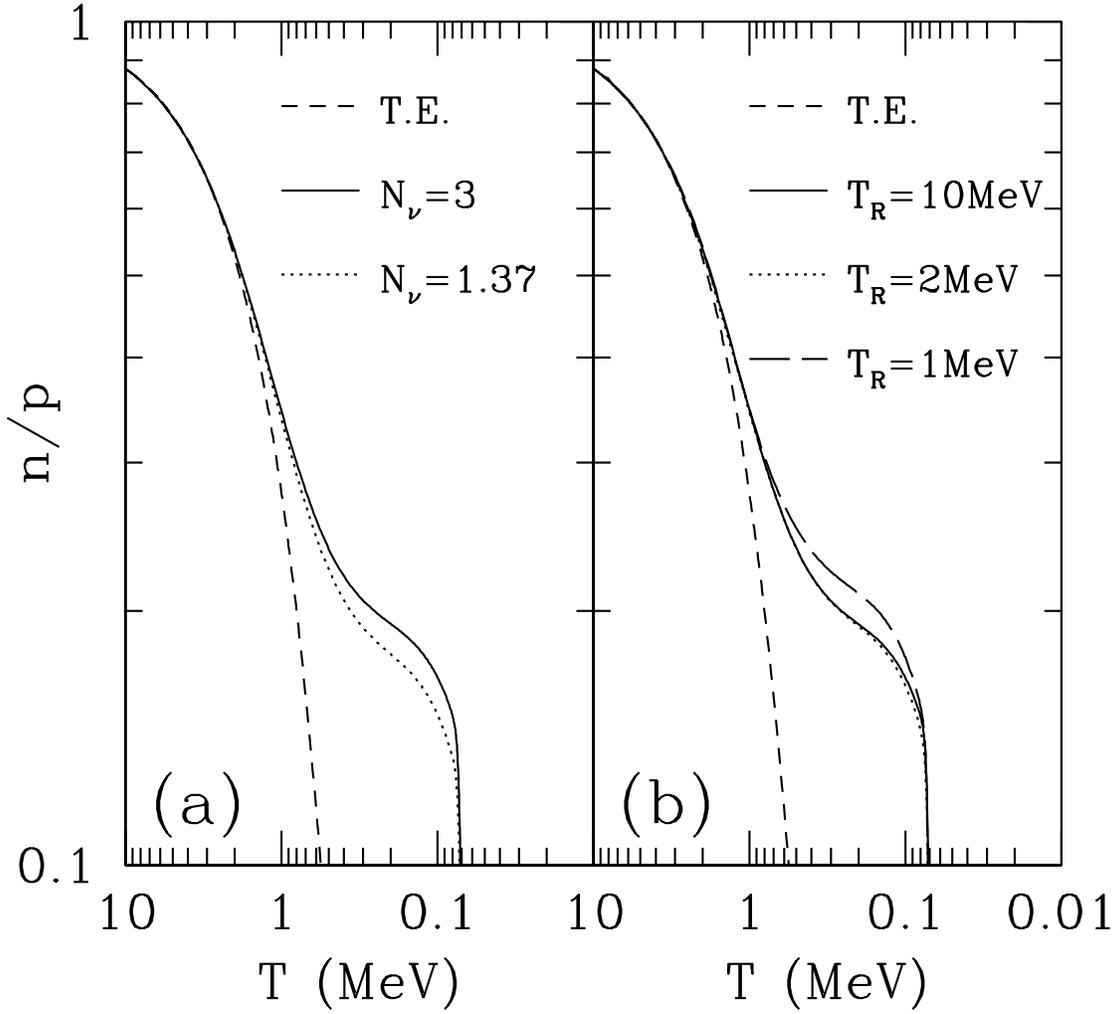,width=16cm}}
       \caption{%%
       Evolution of neutron to proton ratio as a function of the
       temperature, (a)when we change only the number of neutrino
       species in the standard big bang scenario, and (b)when we change
       the reheating temperature in the late-time entropy production
       scenario. The dashed line is the thermal equilibrium curve
       ($= e^{-Q/T}$).  }
       \label{fig:npratio}
   \end{center}
\end{figure}
\newpage
\begin{figure}[htbp]
   \begin{center}
     \centerline{\psfig{figure=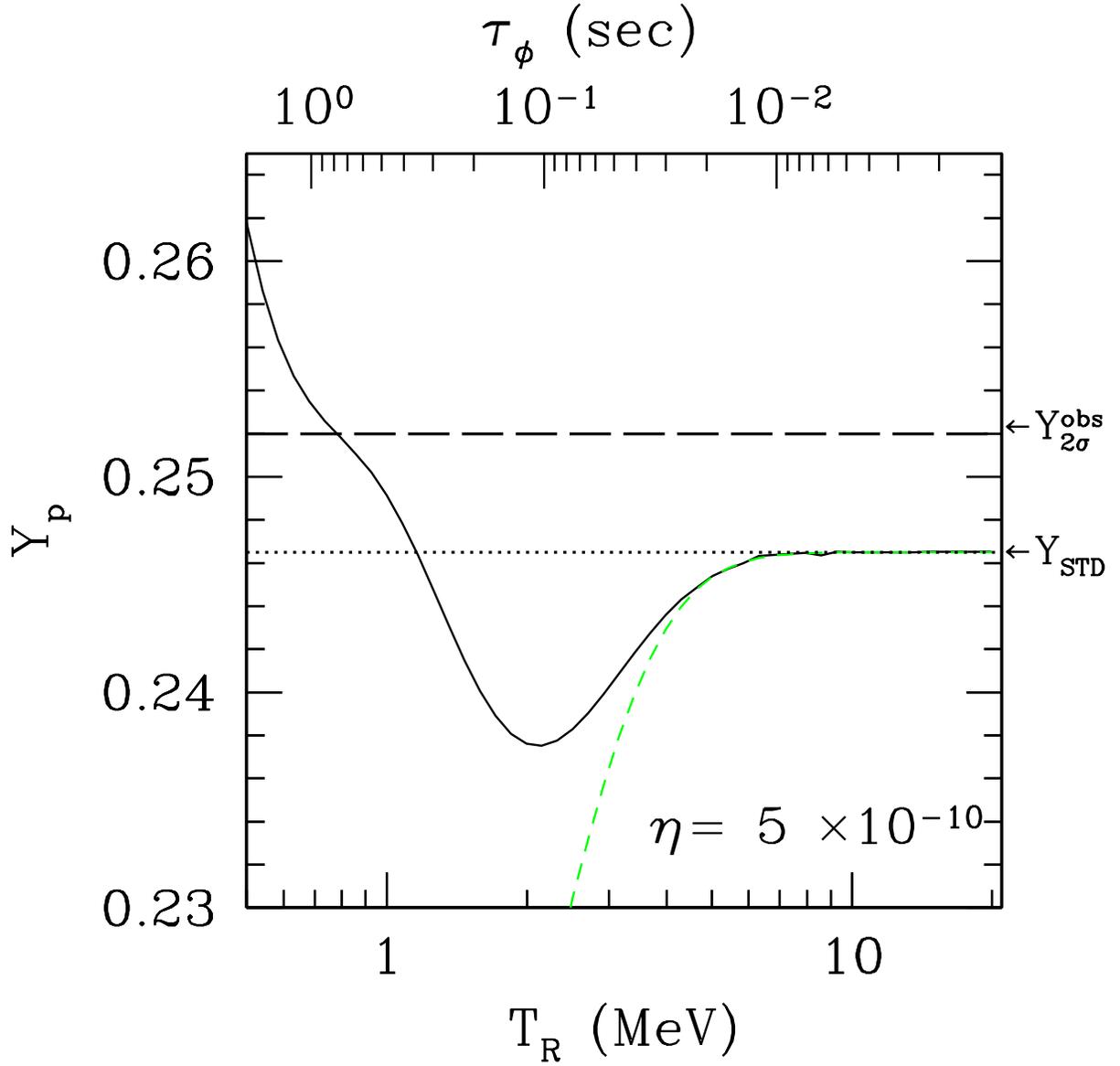,width=18cm}}
       \caption{%%
       $^4$He mass fraction $Y_p$ as a function of $T_R$ (solid line)
       at $\eta = 5 \times 10^{-10}$. The dashed line denotes the
       virtual {}$^4$He mass fraction computed by including only the
       speed down effect due to the change of the effective number of
       neutrino species which is shown in Fig.~\ref{fig:tr_nnu}. The
       dotted line denotes the value predicted in SBBN at $\eta = 5
       \times 10^{-10}$. The long-dashed line denotes the observational
       2 $\sigma$ upper bound, $Y^{obs} \sim 0.252$ which is obtained
       by summing the errors in quadrature. The top horizontal axis
       represents the lifetime which corresponds to $T_R$.}
       \label{fig:pred_he4}
   \end{center}
\end{figure}
\newpage
\begin{figure}[htbp]
   \begin{center}
     \centerline{\psfig{figure=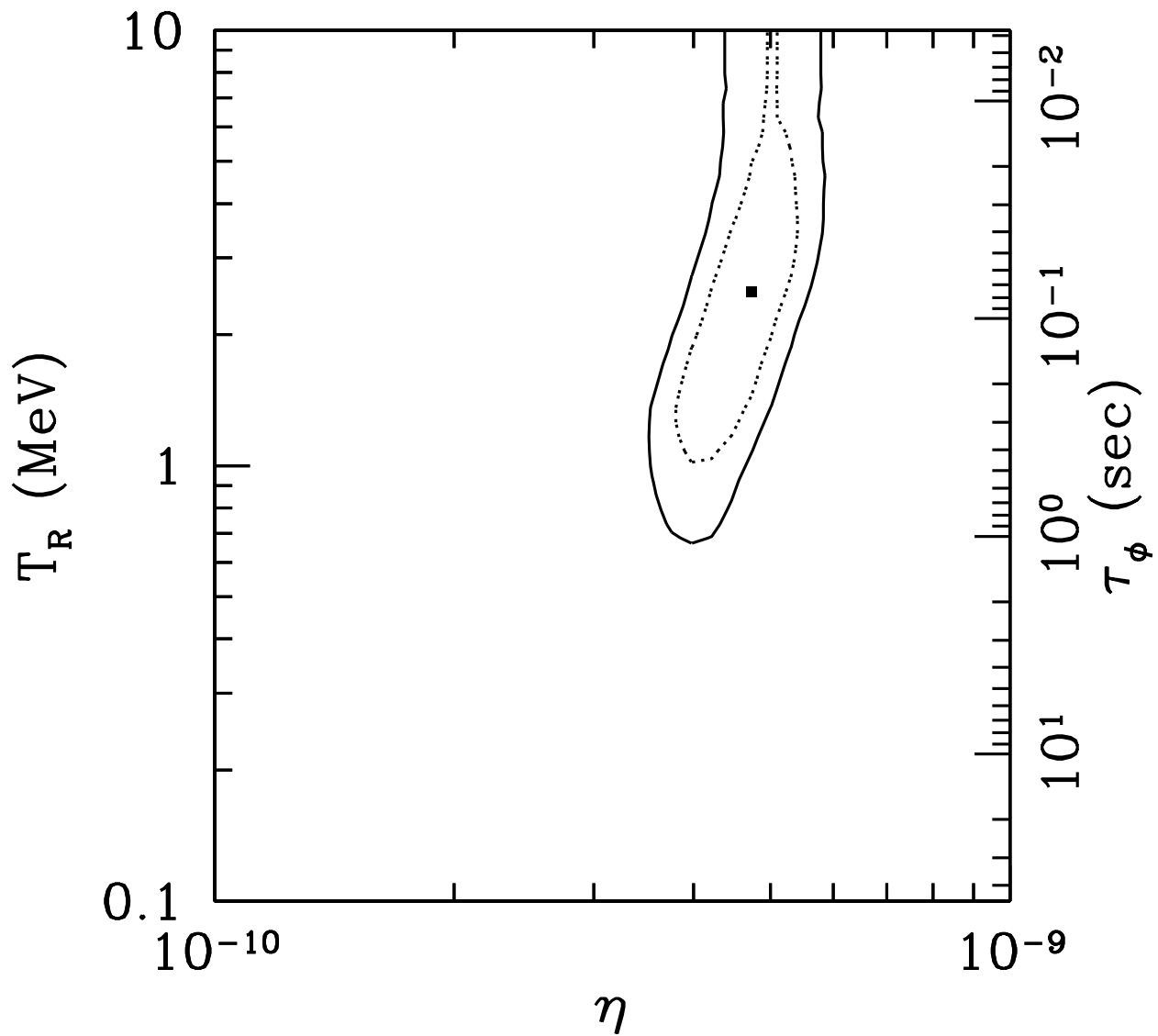,width=18cm}}
       \caption{%%
        Contours of the confidence level in ($\eta,T_R$) plane. The
       inner (outer) curve is  68$\%$ (95$\%$) C.L.. The filled square
       denotes the best fit point. The right vertical
       axis denotes the lifetime which corresponds to
       $T_R$.
       }
       \label{fig:eta_tr}
   \end{center}
\end{figure}
\newpage
\begin{figure}[htbp]
   \begin{center}
     \centerline{\psfig{figure=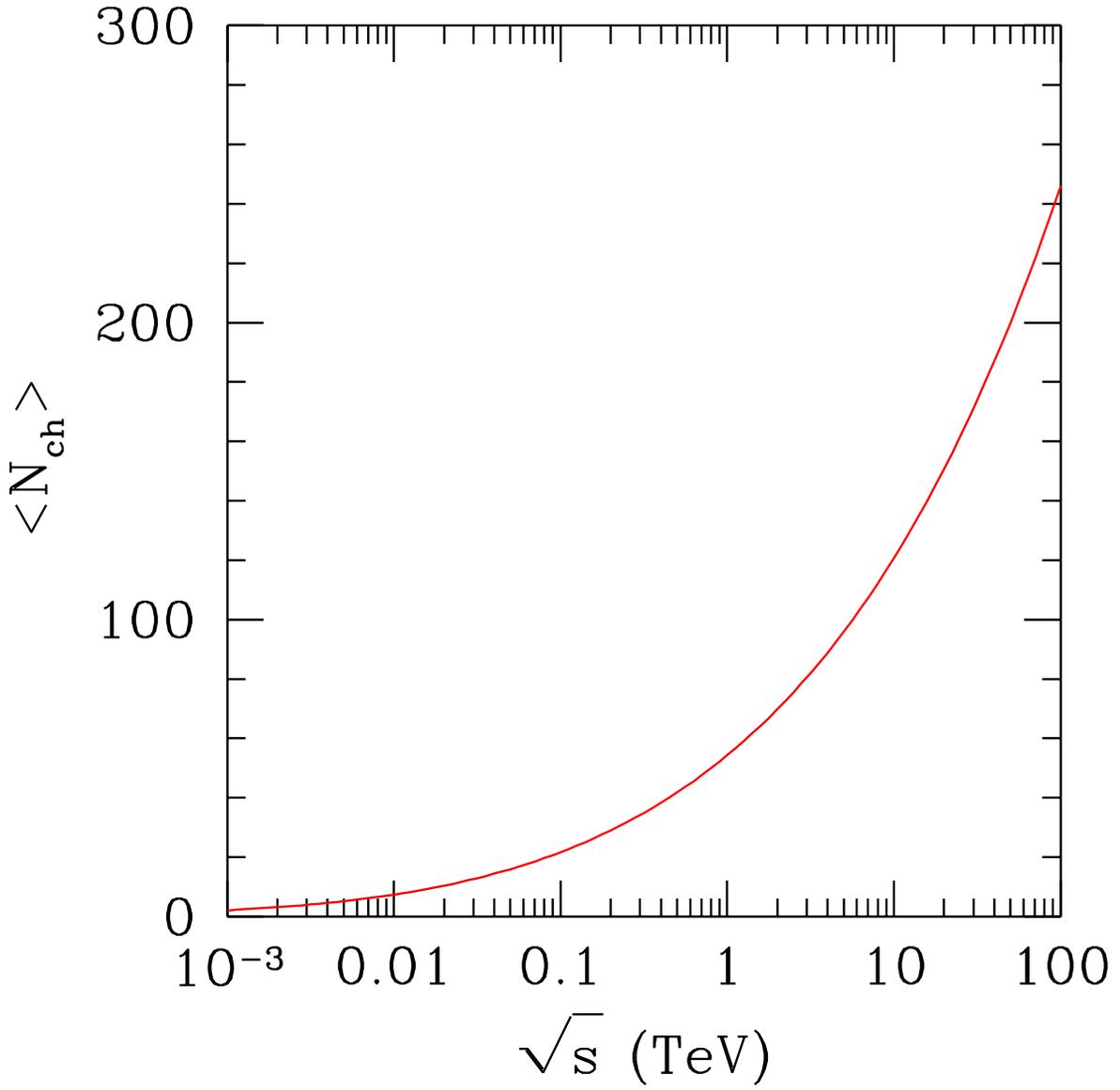,width=18cm}}
       \caption{%%
      Plot of the charged particle multiplicity
       $\langle N_{ch}\rangle$ for the center of mass energy
       $\protect\sqrt{s} = 1 \gev  - 100 \tev$.}
       \label{fig:nch}
   \end{center}
\end{figure}
\newpage
\begin{figure}[htbp]
   \begin{center}
     \centerline{\psfig{figure=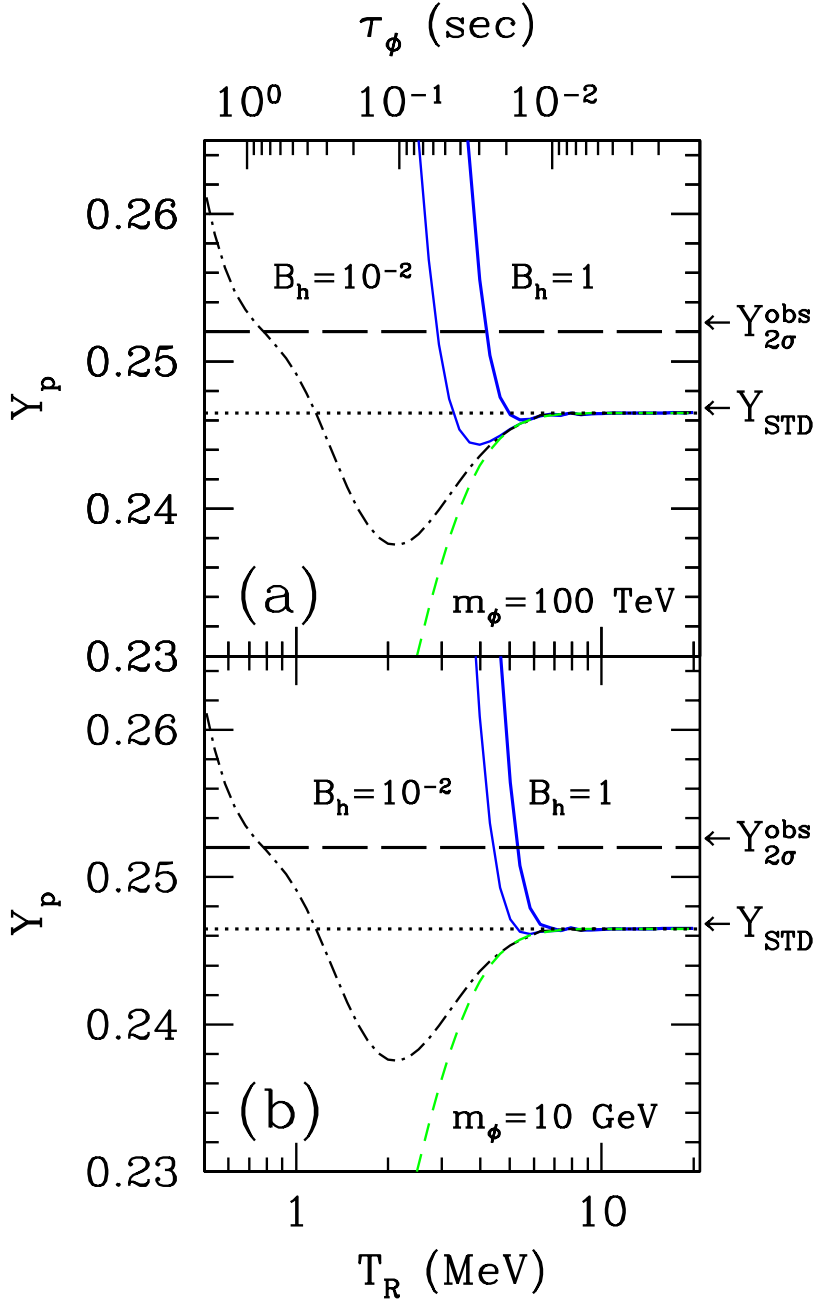,width=18cm}}
       \caption{%%
       Plot of the predicted $\4he$ mass fraction $Y_p$ as a function
       of $T_R$ for (a) m$_{\phi}$=100 TeV and (b) m$_{\phi}$ =10 GeV
       at $\eta = 5 \times 10^{-10}$.  The solid curve denotes the
       predicted $Y_p$ where we take the branching ratio of the
       hadronic decay mode as $B_h$ = 1 (right one) and $B_h$ = 0.01
       (left one). The dot-dashed line denotes $B_h = 0$. The dashed
       line denotes the virtual $Y_p$ curve computed by including only
       the speed down effect due to the change of the effective number
       of neutrino species. The dotted line denotes $Y_p$ in SBBN. The
       long-dashed line denotes the rough observational two $\sigma$
       upper bound that $Y_p$ should be less than about 0.252. The top
       horizontal axis represents the lifetime which corresponds to
       $T_R$.}
       \label{fig:had_tr_Y}
   \end{center}
\end{figure}
\newpage
\begin{figure}[htbp]
   \begin{center}
      \centerline{\psfig{figure=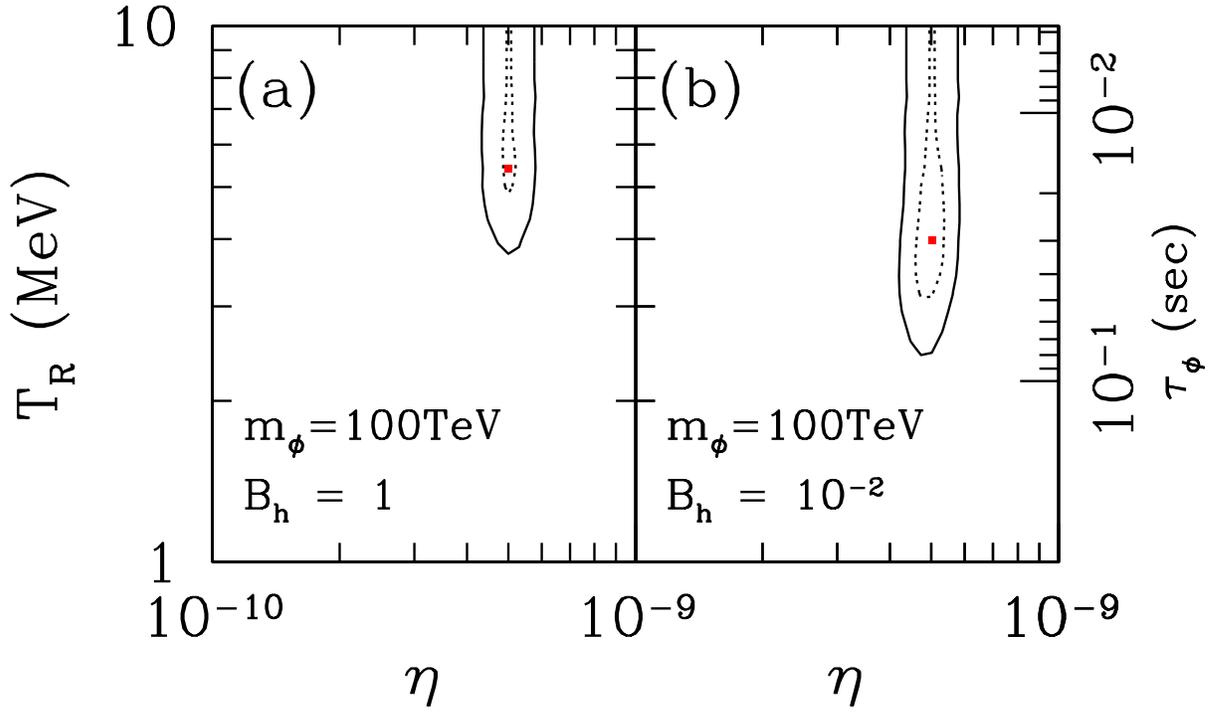,width=18cm}}
      \vspace{-5cm}
       \caption{%%
       Contours of the confidence levels for $m_{\phi}= 100$ TeV in
       ($\eta,T_R$) plane for the branching ratio of the hadronic
       decay mode (a) $B_h$ = 1 and (b) $B_h = 10^{-2}$.  The solid
       line denotes 95 $\%$ C.L. and the dotted line denotes 68 $\%$
       C.L. The filled square is the best fit point between the
       observation and theoretical prediction for D, $\4he$ and $\li7$.
       The right vertical axis represents the lifetime which
       corresponds to $T_R$.}
       \label{fig:m100_eta_tr}
   \end{center}
\end{figure}
\newpage
\begin{figure}[htbp]
   \begin{center}
     \centerline{\psfig{figure=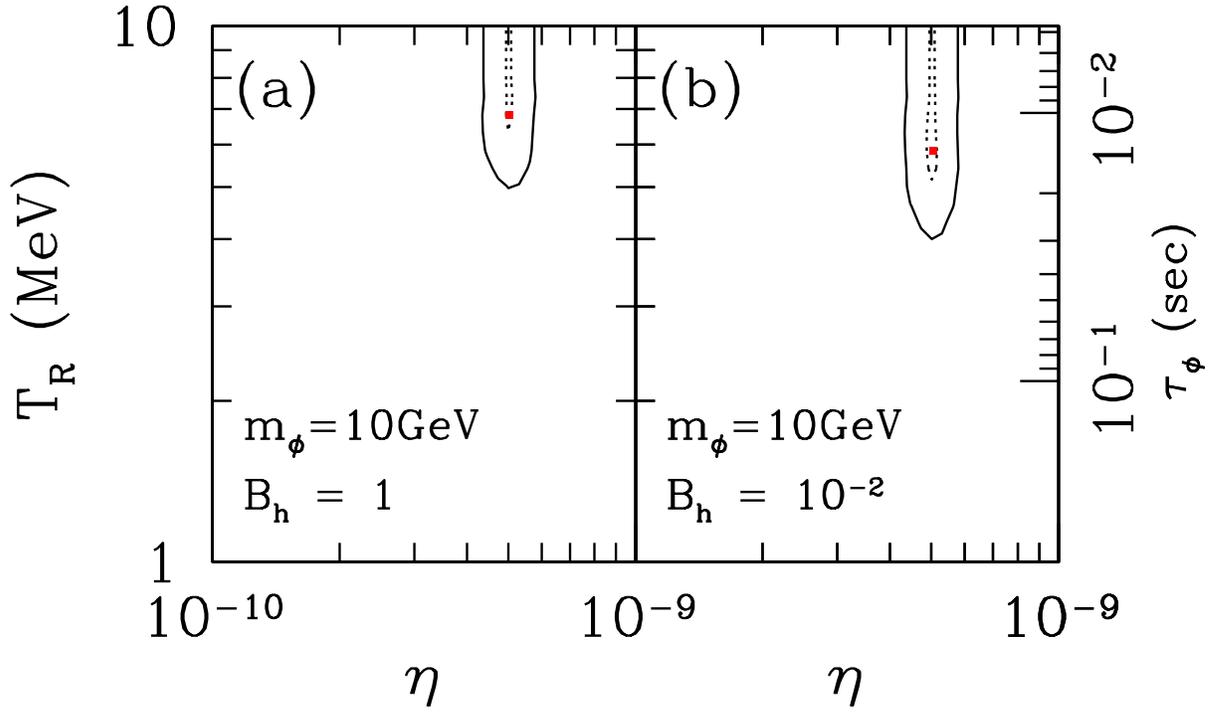,width=18cm}}
     \vspace{-5cm}
       \caption{%%
       Contours of the confidence levels for $m_{\phi}= 10$ GeV for the
       same theory parameters as in Fig.~\ref{fig:m100_eta_tr}. }
       \label{fig:m0.01_eta_tr}
   \end{center}
\end{figure}
\newpage
\begin{figure}[htbp]
   \begin{center}
     \centerline{\psfig{figure=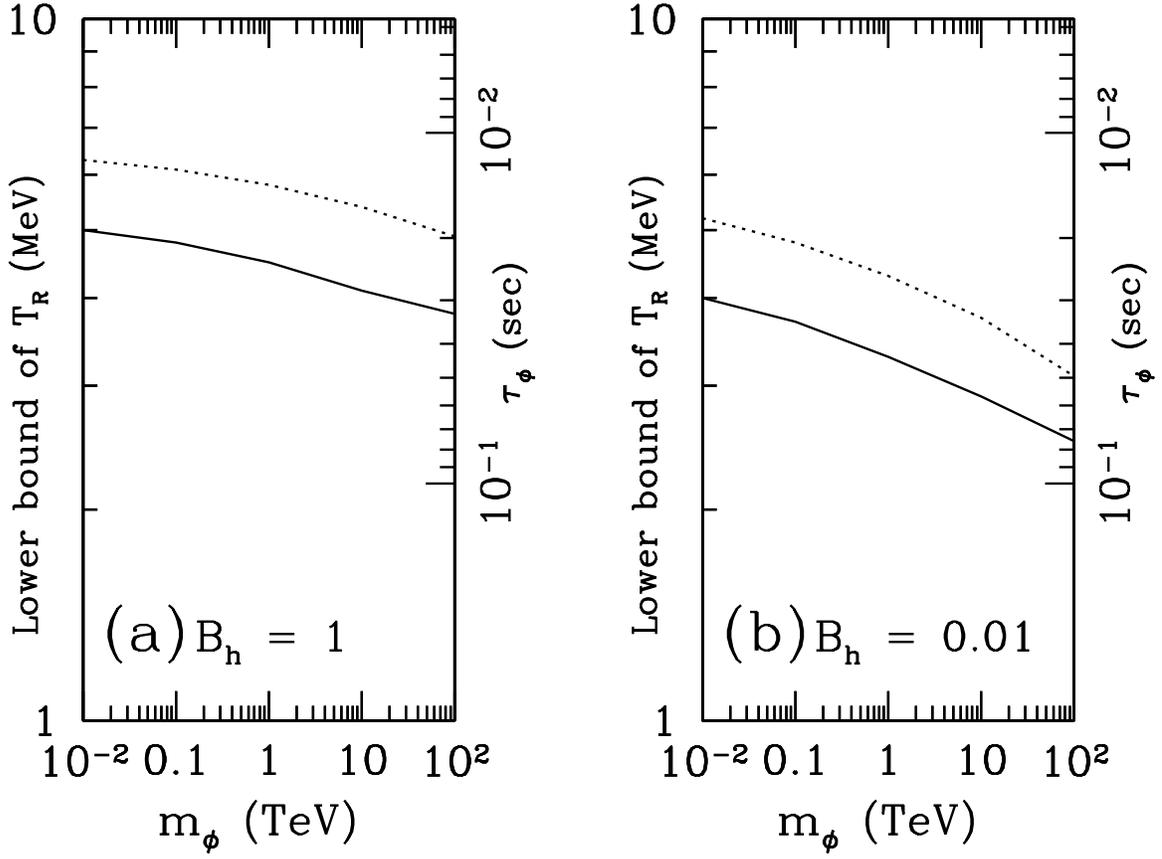,width=18cm}}
     \vspace{-3cm}
       \caption{%%
       Lower bound on $T_R$ as a function of $m_{\phi}$ for the
       branching ratio of the hadronic decay mode (a) $B_h$ = 1 and (b)
       $B_h = 10^{-2}$. The solid line denotes 95 $\%$ C.L.  and the
       dotted line denotes 68 $\%$ C.L. The right vertical axis
       represents the lifetime which corresponds to $T_R$.  }
       \label{fig:m_tr}
   \end{center}
\end{figure}
\newpage
\begin{figure}[htbp]
   \begin{center}
     \centerline{\psfig{figure=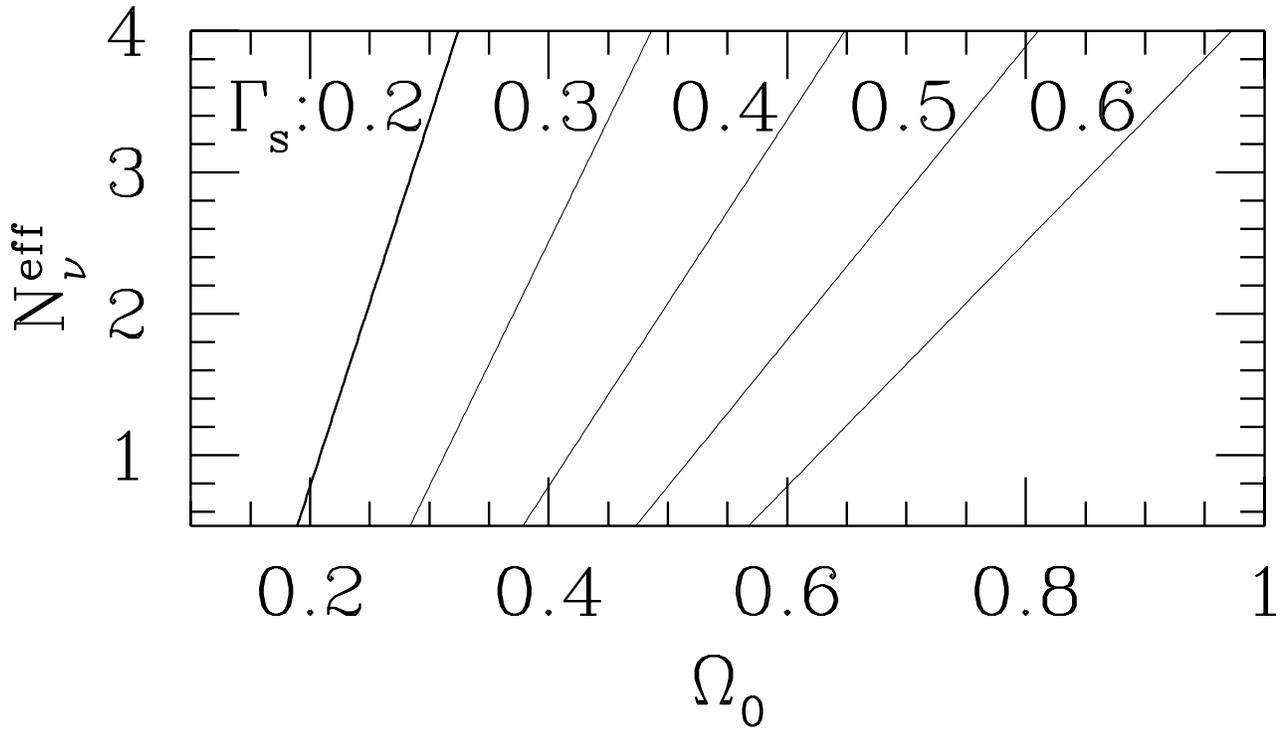,width=18cm}}
%     \vspace{-3cm}
       \caption{%%
       Contours of $\Gamma_{\rm s} = 0.2$ (bold), $0.3, 0.4, 0.5$ and $0.6$ on 
       the ($\Omega_0, N_\nu^{\rm eff}$) plane for $h=0.7$.
        }
       \label{fig:gamma}
   \end{center}
\end{figure}
\begin{figure}[htbp]
   \begin{center}
     \centerline{\psfig{figure=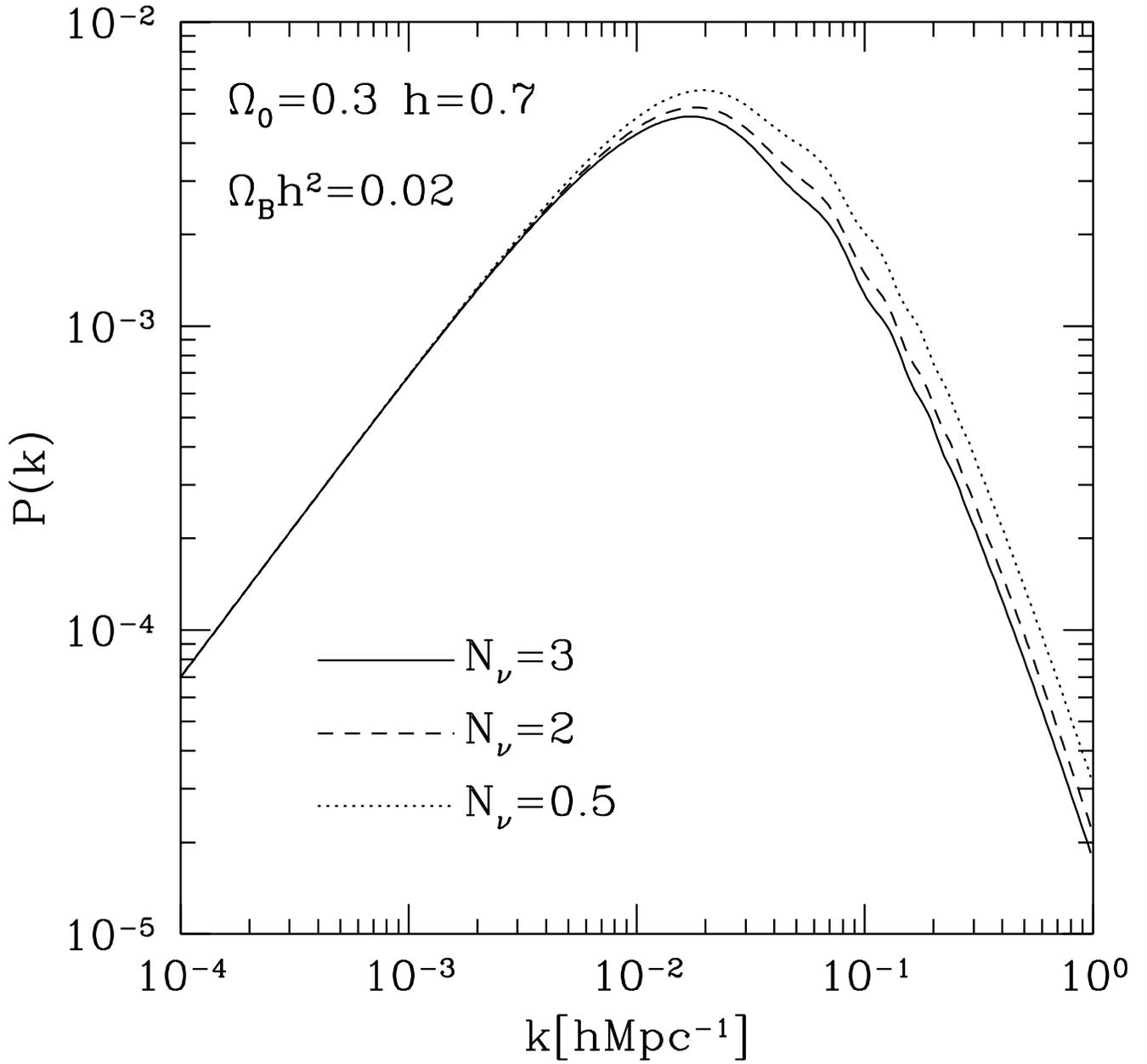,width=18cm}}
       \caption{%%
        Matter power spectra $P(k)$ of CDM models 
        with $N_\nu^{\rm eff}=0.5, 2, $ and $3$.
        We take $\Omega_0=0.3, h=0.7$, and $\Omega_Bh^2=0.02$ where 
        $\Omega_B$ is the baryon density parameter.  
        }
       \label{fig:power}
   \end{center}
\end{figure}
\begin{figure}[htbp]
   \begin{center}
     \centerline{\psfig{figure=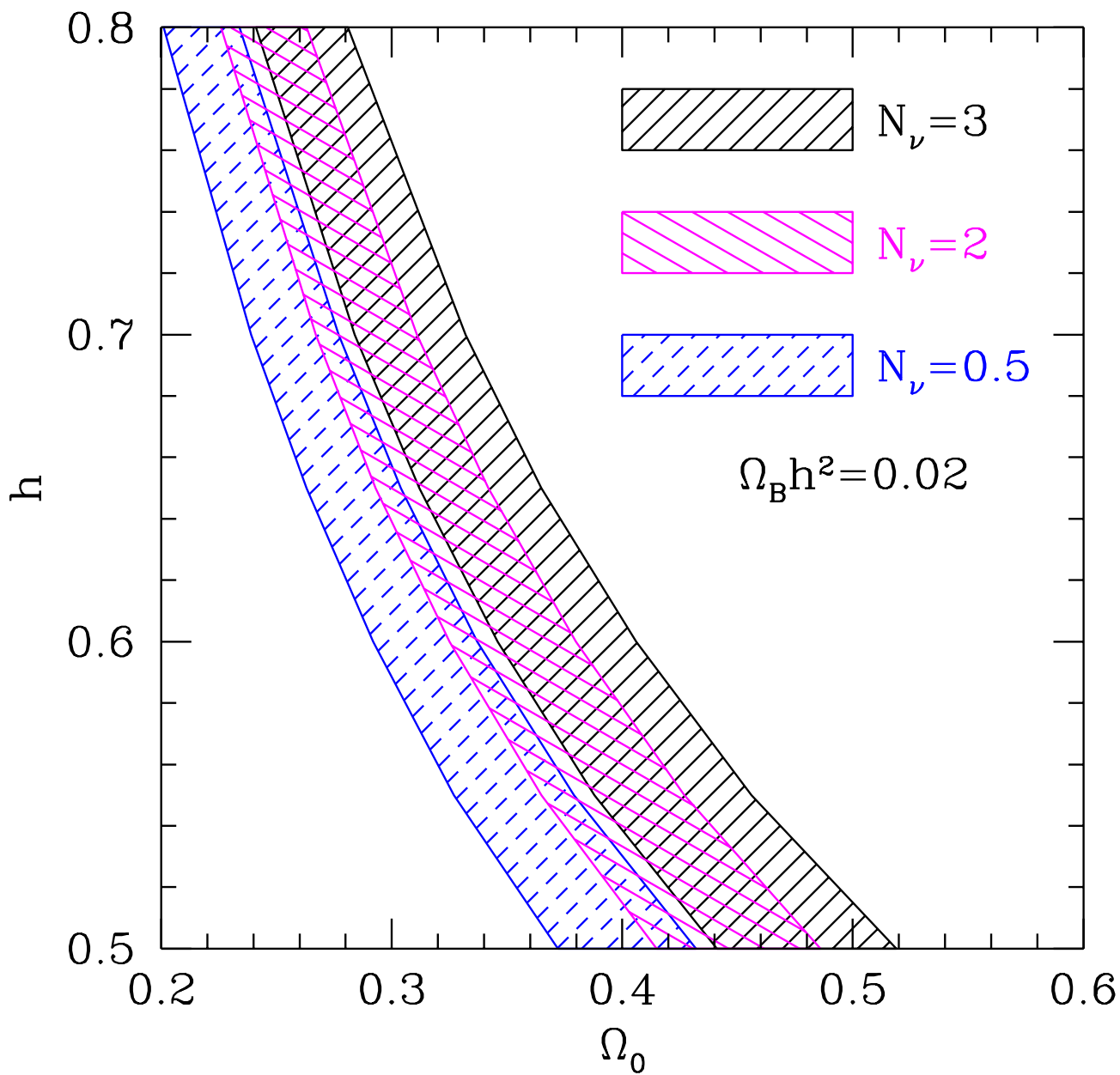,width=18cm}}
%     \vspace{-3cm}
       \caption{%%
        Allowed region on $\Omega_0-h$ plane from observational values of 
        $\sigma_8$ deduced 
        from the rich cluster abundance at present for flat CDM models.
        Models with $N_\nu^{\rm eff}=0.5, 2,$ and $3$ are plotted. 
        }
       \label{fig:sigma8}
   \end{center}
\end{figure}
\begin{figure}[htbp]
   \begin{center}
     \centerline{\psfig{figure=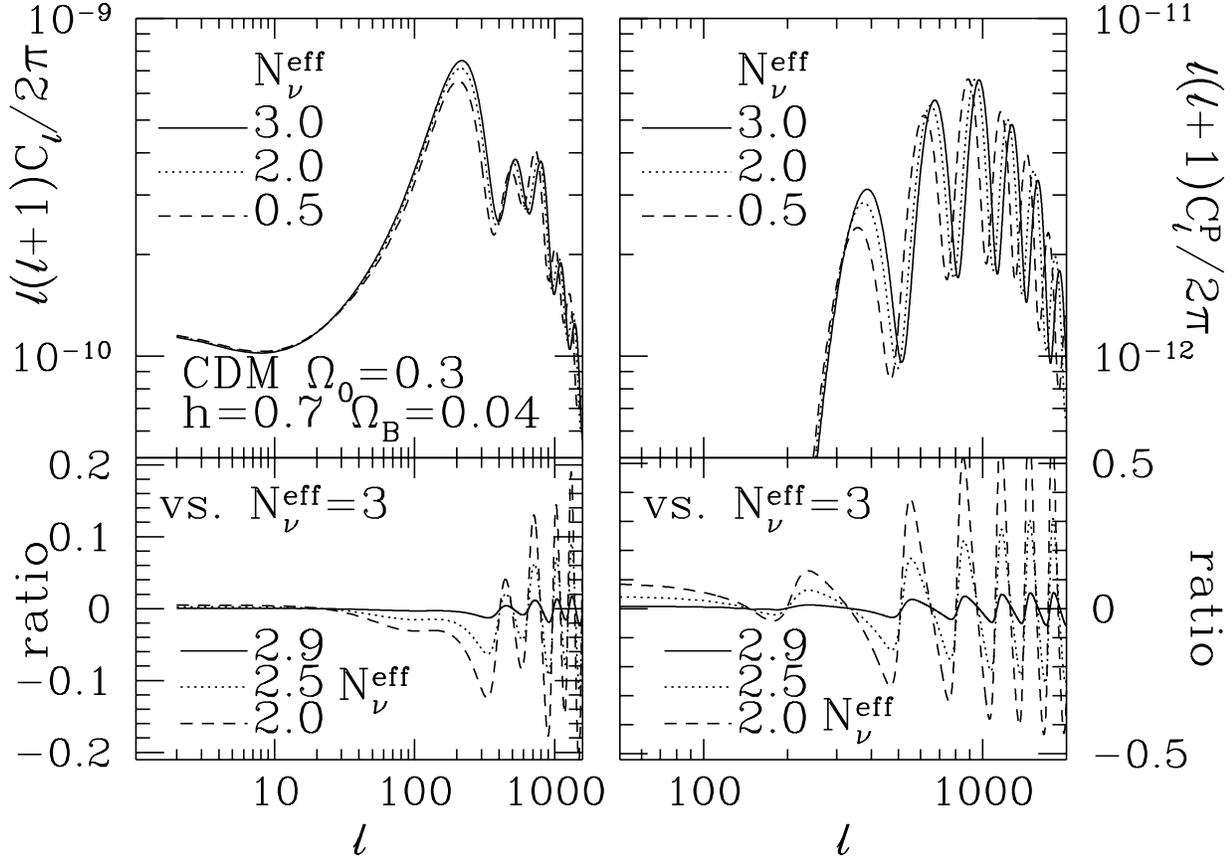,width=16cm}}
%     \centerline{\psfig{figure=cl.ps,width=18cm}}
     \vspace{0.5cm}
%     \vspace{-3cm}
       \caption{%%
        Power spectra of CMB anisotropies (left top panel) and
       polarization (right top panel) of models with 
       $N_{\rm eff}^{\rm eff}=3, 2$
       and $0.5$.  Bottom two panels show $(C_\ell(N_{\rm eff})-
       C_\ell(3))/C_\ell(3)$ with $N_{\rm eff}^{\rm eff}=2.9, 2.5$
       and $2$ for CMB anisotropies (left bottom) and polarization (right 
       bottom).
        }
       \label{fig:CMB}
   \end{center}
\end{figure}

\end{document}